\documentclass[AMA,LATO1COL]{WileyNJD-v2}
 \pdfoutput=1
\articletype{Article Type}%

\received{26 April 2016}
\revised{6 June 2016}
\accepted{6 June 2016}
\usepackage{comment}
\usepackage{pgfplots, pgfplotstable,booktabs}
\usepackage{tcolorbox}
\usepackage{pgf-pie}
\usepackage{tikz}
\usetikzlibrary{tikzmark}
\usetikzlibrary{fit} 
\usetikzlibrary{positioning}
\usetikzlibrary{arrows}
\usetikzlibrary{shapes.multipart}
\usepackage{pstricks}
\usetikzlibrary{patterns,}
\usepackage{amsmath}
\usepackage{subcaption}
\usepackage{dirtytalk}
\pgfplotsset{compat=1.14}
\definecolor{findOptimalPartition}{HTML}{696969}
\definecolor{storeClusterComponent}{HTML}{808080}
\definecolor{dbscan}{HTML}{BEBEBE}
\definecolor{constructCluster}{HTML}{DCDCDC}

\newcommand{\RQA}{\textbf{RQ1.} What  is  the  distribution  of  experience  among  developers  that  perform refactorings?}
\newcommand{\RQB}{\textbf{RQ2.} Do developers with more contribution refactor code more often?}
\newcommand{\RQC}{\textbf{RQ3.} What triggers developers to refactor the code?}
\newcommand{\RQD}{\textbf{RQ4.} Does developer's experience influence the quality of refactoring documentation?}

\pgfplotstableread[col sep=comma,header=true]{
Category,1,2,3,4,5
Change Package, 12.1, 17.0, 14.7, 18.6, 37.6
Extract \& Move Method, 14.7, 28.3, 21.1, 12.4, 13.6
Extract Class, 18.6, 25.1, 23.1, 14.3, 19.0
Extract Interface, 20.5, 22.3, 19.8, 14.9, 22.5
Extract Method, 28.7, 24.5, 19.1, 14.7, 13.0
Extract Subclass, 17.7, 21.9, 29.3, 11.9, 19.3
Extract Superclass, 20.7, 25.5, 19.6, 17.1, 17.1
Extract Variable, 35.7, 21.7, 17.2, 16.1, 9.3
Inline Method, 28.7, 20.1, 19.9, 15.8, 15.6
Inline Variable, 30.4, 18.3, 20.5, 16.5, 14.4
Move \& Rename Attribute, 21.2, 32.7, 26.9, 13.5, 5.8
Move \& Rename Class, 12.4, 20.1, 19.5, 20.8, 27.2
Move Attribute, 19.1, 21.6, 26.6, 12.0, 20.7
Move Class, 9.9, 15.1, 15.1, 13.7, 46.1
Move Method, 17.4, 24.8, 20.7, 11.6, 25.5
Move Source Folder, 15.5, 20.1, 18.7, 27.0, 18.7
Parameterize Variable, 25.9, 26.0, 18.8, 16.7, 12.7
Pull Up Attribute, 21.9, 19.6, 19.6, 13.8, 25.1
Pull Up Method, 21.4, 21.0, 23.0, 17.1, 17.5
Push Down Attribute, 14.8, 20.7, 26.2, 17.0, 21.3
Push Down Method, 13.3, 17.4, 28.4, 16.0, 24.9
Rename Attribute, 23.2, 18.5, 17.5, 19.2, 21.6
Rename Class, 20.6, 19.1, 17.6, 19.2, 23.5
Rename Method, 22.1, 20.9, 18.3, 19.3, 19.4
Rename Parameter, 24.6, 22.6, 23.1, 14.8, 14.9
Rename Variable, 22.9, 22.8, 24.2, 15.2, 14.8
Replace Attribute, 15.4, 15.4, 48.7, 7.7, 12.8
Replace Variable W/ Attribute, 27.9, 24.9, 21.3, 14.3, 11.6
}\dataA

\pgfplotstablecreatecol[
 create col/expr={
    \thisrow{1} + \thisrow{2} + \thisrow{3} +\thisrow{4} +\thisrow{5}
 }
]{sum}{\dataA}

\pgfplotsset{
  percentage plot/.style={
    point meta=explicit,
    every node near coord/.append style={
      font=\tiny,
      color=black,
    },
    nodes near coords={
      \pgfmathtruncatemacro\iszero{\originalvalue==0}
      \ifnum\iszero=0
      \pgfmathprintnumber[fixed,fixed zerofill,precision=1]{\pgfplotspointmeta}
      \fi
    },
    yticklabel=\pgfmathprintnumber{\tick}\,$\%$,
    ymin=0,
    ymax=100.01, 
    visualization depends on={y \as \originalvalue},
    enlarge x limits={abs=10mm}
  },
  percentage series/.style={
    table/x expr=\coordindex, 
    table/y expr=(\thisrow{#1}/\thisrow{sum}*100),
    table/meta=#1
    }
}

\begin{document}


\title{Behind the Scenes: On the Relationship Between Developer Experience and Refactoring}

\author[1]{Eman Abdullah AlOmar*}

\author[2]{Anthony Peruma}

\author[3]{Mohamed Wiem Mkaouer}

\author[4]{Christian D. Newman}

\author[5]{Ali Ouni}

\authormark{AlOmar \textsc{et al}}

\address[1]{\orgdiv{Department of Software Engineering}, \orgname{Rochester Institute of Technology}, \orgaddress{\state{New York}, \country{USA}}}

\address[2]{\orgdiv{Department of Software Engineering}, \orgname{Rochester Institute of Technology}, \orgaddress{\state{New York}, \country{USA}}}

\address[3]{\orgdiv{Department of Software Engineering}, \orgname{Rochester Institute of Technology}, \orgaddress{\state{New York}, \country{USA}}}

\address[4]{\orgdiv{Department of Software Engineering}, \orgname{Rochester Institute of Technology}, \orgaddress{\state{New York}, \country{USA}}}

\address[5]{\orgdiv{Department of Software Engineering and IT}, \orgname{ETS Montreal}, \orgaddress{\state{Montreal}, \country{Canada}}}

\corres{*Eman Abdullah AlOmar. \email{eman.alomar@mail.rit.edu}}


\abstract[Summary]{
Refactoring is widely recognized as one of the efficient techniques to manage technical debt and maintain a healthy software project through enforcing best design practices, or coping with design defects. Previous refactoring surveys have shown that code refactoring activities are mainly executed by developers who have sufficient knowledge of the system's design, and disposing of leadership roles in their development teams. However, these surveys were mainly limited to specific projects and companies. In this paper, we explore the generalizability of the previous results by analyzing 800 open-source projects. We mine their refactoring activities, and we identify their corresponding contributors. Then, we associate an experience score to each contributor in order to test various hypotheses related to whether developers with higher scores tend to 1) perform a higher number of refactoring operations 2) exhibit different motivations behind their refactoring, and 3) better document their refactoring activity. We found that (1) although refactoring is not restricted to a subset of developers, those with higher contribution score tend to perform more refactorings than others; (2) while there is no correlation between experience and motivation behind refactoring, top contributed developers are found to perform a wider variety of refactoring operations, regardless of their complexity; and (3) top contributed developer tend to document less their refactoring activity. Our qualitative analysis of three randomly sampled projects show that the developers who are responsible for the majority of refactoring activities are typically in advanced positions in their development teams, demonstrating their extensive knowledge of the design of the systems they contribute to.}


\maketitle



\section{Introduction} \label{section:introduction}



According to the Consortium for Information \& Software Quality, poor software quality costs the United States economy over 2 trillion, in 2020, due to functional software failures, poor quality of existing legacy systems, and unsuccessful projects delivery \footnote{\url{https://www.it-cisq.org/the-cost-of-poor-software-quality-in-the-us-a-2020-report.htm}}. Therefore, refactoring was born along with code reviews, as a natural response to be the quality safeguard and the backbone of managing technical debt \cite{codabux2013managing}. The main goal of refactoring is to restructure the design and the source code to be more efficient and easier to comprehend. It is key in reducing the cost maintaining and evolving software, as it makes the process of debugging smoother. While it is not intended as a bug fix practice, it has been found to reduce software proneness to defects \cite{bavota2015experimental}.

The spectrum of research exploring the practice of refactoring covers a wide variety of dimensions, such as the identification of refactoring opportunities \cite{fontana2012automatic,palomba2013detecting,palomba2016textual}, recommendation of adequate refactoring operations \citep{bavota2014recommending,mkaouer2014high,mkaouer2015many,mkaouer2017robust,mkaouer2014software,rizzi2018support,terra2018jmove,de2019finding}, detection of applied refactorings \cite{xing2006refactoring,kim2010ref,silva2017refdiff}, studying the impact of refactoring on quality \cite{xing2006eclipse,bavota2015experimental,pinto2013programmers,yoshida2016revisiting}, the reasons as to why developers refactor their code \cite{silva2016we,peruma2018empirical}, etc. However, little is known about how the level of experience influences developer refactoring activities. Nevertheless, developer experience directly impacts their ability to estimate software quality, and therefore, their ability to determine the appropriate refactoring strategy that needs to be deployed. Moreover, developers' knowledge of the system's structure and sub-components varies, and so is their privilege to access and modify them. This paper aims to start the discussion around the importance of considering the developer's experience as part of proposing solutions related to refactoring, since their applicability depends on the perception and privilege of the developers in charge. 

In our previous work \cite{alomar2020relationship}, we explored the hypothesis of whether developers with more contribution are most likely to be responsible for a higher number of refactoring activities. This study extends our prior investigation by exploring two more hypotheses: 1) We argue whether developers with higher contribution are most likely to have different motivations to refactor code; and 2) we investigate whether developers top contributors correlates with better documentation of refactorings. Our first hypothesis is driven by the assumption that top contributed developers are most likely to perform more complex refactorings. Our second hypothesis argues that top contributed developers may provide clearer documentation of their code changes, as a reflection of better understanding the value of proper documentation. 
Our study is driven by the following research questions:
\begin{itemize}
\item \RQA

To answer this research question, we start with mining refactorings from 800 well-engineered projects. We identify the subset of authors who were involved in these refactoring activities along with all project contributors. We estimate their contribution in the project by measuring their developer commit ratio score. We compare the scores of developers whose commits witnessed refactorings with the scores of developers whose commits had no refactorings.

\item \RQB 

The rationale behind this question is investigating whether refactoring activities tend to be performed by a subset of developers. To answer this question, we split developers, based on their experience score, into two sets, where the first set contains the top 5\% of developers with high contribution scores while the second set gathers the remaining contributors. Then we compared the count of the refactorings performed by each set. We further randomly sampled three projects, and we extracted their top contributors with respect to refactoring both production and test code.


\item {\RQC} 

This research question investigates what motivates developers to refactor their code, by identifying the type of development tasks in which refactorings were interleaved, e.g., updating a feature, debugging, etc. We verify if developer's experience correlates with a specific motivation. We also breakdown our analysis per type of refactoring operation performed. 


\item {\RQD 
}

Answering this question helps to explore what type of refactoring contributors frequently document refactoring activities in their commit messages, and whether experience plays a role in providing a better documentation of performed refactoring operations.
\end{itemize}


The remainder of this paper is organized as follows. Section \ref{section:related work} discusses the related work. Section \ref{section:methodology} outlines our experimental methodology in collecting the necessary refactoring data for the experiments that are discussed afterward in Section \ref{section:results}. The research implication is discussed in Section \ref{section:implication}. Section \ref{section:threats} gathers potential limitations to the validity of our empirical analysis before concluding and describing our future work directions in Section \ref{sec:conclusion}.

\section{Related Work} \label{section:related work}

We divided the related work section into two areas– studies that investigated the relationship of refactoring and developer experience, and studies that investigated refactoring documentation in the commit messages.

\subsection{Refactoring \& Developer Experience}
\begin{table*}[h!]
\begin{center}
\caption{Existing works on refactoring \& developer experience}
\label{Table:Related Work-refactoring experience}
 \begin{tabular}{ llllll } \hline
  \toprule
  \bfseries Study  & \bfseries Year & \bfseries Subject & \bfseries Approach  & \bfseries Source of Info. & \bfseries Main Finding \\
  \midrule
    Tsantalis et al. \cite{tsantalis2013multidimensional} & 2013 & 3 OSS &  Manual Analysis & Refactoring Commits &  Refactoring contributors has the management \\
    & & & & & role  within the project \\
    Kim et al. \cite{kim2014empirical} & 2014 & 328 developers & Survey &  Refactoring Commits & Developer experience needs to be examined \\
    & & & & & as it might be the cause for changes of module \\
    & & & & & dependencies   \\
    AlOmar et al. \cite{alomar2020relationship} & 2020  & 800 OSS & DCR Calculation & Refactoring Commits & Higher experience developers perform the  \\
    & & & & & majority of refactoring \\
  \bottomrule
  \end{tabular}
  \end{center}
\end{table*}
A couple of refactoring studies have pointed out that refactoring is typically performed by experienced developers: Tsantalis et al. \cite{tsantalis2013multidimensional} performed a multidimensional empirical study on refactoring activities that included: the proportion of refactoring operations performed on production and test code, the most active refactoring contributors, the relationship between refactorings with releases and testing activity, and the purpose of the applied refactorings. With regard to developer experience, the authors found that the top refactoring contributors had a management role within the project. In another study, Kim et al. \cite{kim2014empirical} surveyed 328 professional software engineers at Microsoft to investigate when and how they do refactoring. They found that developers with different expertise levels experienced five risk factors involved in refactoring, namely, regression bugs, code churns, merge conflicts, time taken from other tasks, the difficulty of performing code reviews after refactoring, and the testing cost. They also investigated the relationship between the refactoring effort and reduction of the number of inter-module dependencies and after release defects. They reported that other factors, such as developer experience, need to be examined as the changes to the number of module dependencies and post-release defects might be caused by such factors other than refactoring. The findings of these studies \cite{tsantalis2013multidimensional,kim2014empirical,newman2018study} indicate that experience plays a significant role in the execution of refactoring, yet they were both limited to developer surveys without any concrete evidence from the source code, and they were also limited to a few projects. More recently, AlOmar et al. \cite{alomar2020relationship} explored the generalizability of Tsantalis et al. study \cite{tsantalis2013multidimensional} by analyzing 800 open-source projects, finding that developers with higher contribution perform the majority of refactoring activities than others. A summary of the related studies is provided in Table \ref{Table:Related Work-refactoring experience}.

\subsection{Refactoring Documentation}
\begin{table*}[h!]
\begin{center}
\caption{Existing works on refactoring identification}
\label{Table:Related Work-Refactoring Identification}
\resizebox{\textwidth}{!}{%
 \begin{tabular}{ llllll } \hline
  \toprule
  \bfseries Study  & \bfseries Year & \bfseries Purpose & \bfseries Approach  & \bfseries Source of Info. & \bfseries Ref. Patterns\\
  \midrule
   Stroggylos \& Spinellis  \cite{stroggylos2007refactoring} & 2007 & Identify refactoring commits &  Mining commit logs & General commits & 1 keyword \\
   Ratzinger et al. \cite{Ratzinger:2008:RRS:1370750.1370759} & 2007 \& 2008 & Identify refactoring commits &  Mining commit logs & General commits & 13 keywords  \\
   Murphy-Hill et al.  \cite{murphy2012we} & 2012 & Identify refactoring commits & Ratzinger's approach & General commits & 13 keywords \\
   Soares et al. \cite{soares2013comparing} & 2013 & Analyze refactoring activity & Ratzinger's approach & General commits & 13 keywords \\
    & & & Manual analysis & & \\
    & & & Dynamic analysis & & \\
    Kim et al. \cite{kim2014empirical} & 2014 & Identify refactoring commits & Identifying refactoring branches &  Refactoring branch & Top 10 keywords\\
    & &  &   Mining commit logs \\
     Zhang et al. \cite{zhangpreliminary18} & 2018 & Identify refactoring commits & Mining commit logs & General commits & 22 keywords \\
     AlOmar et al. \cite{alomar2019can,alomar2019towards,alomar2020howwe} & 2019 \& 2020 & Identify refactoring patterns & Detecting refactorings & Refactoring commits &    87 \& 513 keywords \& phrases   \\
    & & & Extracting commit messages & & \\
  \bottomrule
  \end{tabular}}
  \end{center}
\end{table*}

As shown in Table \ref{Table:Related Work-Refactoring Identification}, a line of works have focused on identifying refactoring activities via commit messages analysis. Stroggylos \& Spinellis \cite{stroggylos2007refactoring} searched for the term `refactor' to study refactoring-related commits. Ratzinger et al.  \cite{Ratzinger:2008:RRS:1370750.1370759} also opted for a similar keyword-based approach to detect refactoring activity in the commit messages. They identified the following 13 terms in their search approach: `refactor', `restruct', `clean', `not used', `unused', `reformat', `import', `remove', `replace', `split', `reorg', `rename', and `move'. Later, Murphy-Hill et al. \cite{murphy2012we} replicated Ratzinger's experiment in two open source systems using Ratzinger's 13 keywords. They conclude that commit messages are unreliable indicators of refactoring activities. This is due to the fact that developers do not consistently report/document refactoring activities in the commit messages. Soares et al. \cite{soares2013comparing} compared three approaches, namely,  manual analysis, commit message (Ratzinger et al.'s approach \cite{Ratzinger:2008:RRS:1370750.1370759}), and dynamic analysis (SafeRefactor approach \cite{Soares2009safetytool}) to analyze refactorings in open source repositories, in terms of behavioral preservation. Their findings show that manual analysis achieves the best results in this comparative study and is considered as the most reliable approach in detecting behavior-preserving transformations. In an industrial survey involving 328 developers at Microsoft, Kim et al. \cite{kim2014empirical} surveyed professional software engineers to investigate when and how they do refactoring. They first identified refactoring branches and then asked developers about the keywords that are usually used to mark refactoring events in change commit messages. When surveyed, the developers mentioned several keywords to mark refactoring activities. Kim et al. matched the top ten refactoring-related keywords identified from the survey (`refactor', `clean-up', `rewrite', `restructure', `redesign', `move', `extract', `improve', `split', `reorganize', `rename') against the commit messages to detect refactoring commits. Using this approach, they found 94.29\% of commits do not have any of the keywords, and only 5.76\% of commits included refactoring-related keywords. Zhang et al. \cite{zhangpreliminary18} performed a preliminary investigation of Self-Admitted Refactoring (SAR) in three open source systems. They first extracted 22 keywords from a list of refactoring operations defined in the Fowler's book \cite{Fowler:1999:RID:311424} as a basis for SAR identification.

AlOmar et al. \cite{alomar2019can,alomar2020howwe} performed an exploratory study on how developers document their refactoring activities in commit messages; this activity is called Self-Affirmed Refactoring (SAR). They found that developers tend to use a variety of textual patterns to document their refactoring activities, such as \textit{refactor}, \textit{move} and \textit{extract}. Since the manual extraction of refactoring patterns is a human intensive task and it is subject
to personal bias, a refactoring model is built to automate the identification of refactorings based on refactoring patterns and their
quality improvement categories \cite{alomar2020toward}. In a subsequent study \cite{alomar2019impact}, the authors identified which quality models are more in-line with the developer's vision of quality optimization when they explicitly mention in the commit messages that they refactor to improve these quality attributes. In their study of the motivation behind the application of refactoring, AlOmar et al. \cite{alomar2020howwe} text-mined refactoring-related documentation and automatically classify a large set of commits containing refactoring activities. The authors proved the existence of motivations that go beyond the basic need for improving the system’s design as the main drivers for refactoring activities include the following categories: Functional, Bug Fix, Internal Quality Attribute, Code Smell Resolution, and External Quality Attribute.

Since we noticed in our previous work \cite{alomar2020relationship} that various developers are responsible for performing refactorings, in this follow-up work, we explore what triggers expert developers to refactor the code and what is their refactoring documentation practices. Unlike our study, prior works merely identified the relationship between expertise and refactoring in general without taking developer perception and documentation into consideration. Since the investigation of the relationship between refactoring activities and developer experiences is important to understand the practice of refactoring, in this paper, we push research on refactoring documentation a step forward by exploring which developers are responsible for the introduction of refactoring patterns in order to examine whether or not experience plays a role in the introduction of these patterns, performed automatically over a much larger sample of commit messages.

\section{Study Design} \label{section:methodology}
\begin{figure}[]
 	\centering
 	\includegraphics[width=1.0\linewidth]{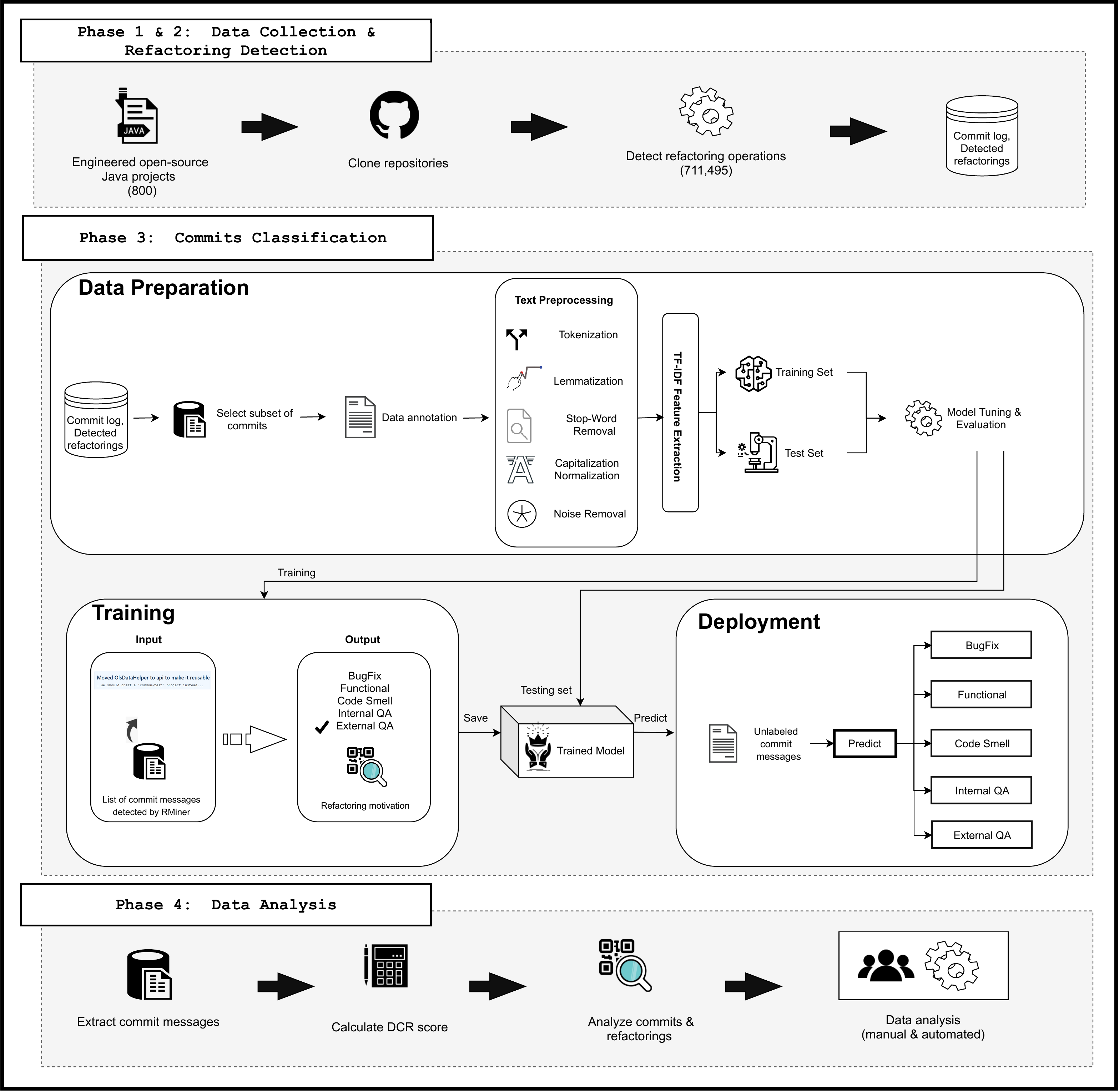}
 	\caption{Overview of our methodology}
 	\label{Figure:methodology}
\end{figure}

Our research methodology consists of three main phases - Data Collection, Refactoring Detection \& Extraction, and Data Analysis. Figure~\ref{Figure:methodology} provides an overview of our methodology. Described below are details of the methodology activities.

\subsection{Data Collection}

\begin{figure}[t]
 \centering
\begin{subfigure}{0.3\linewidth}
 \includegraphics[width=1.\linewidth]{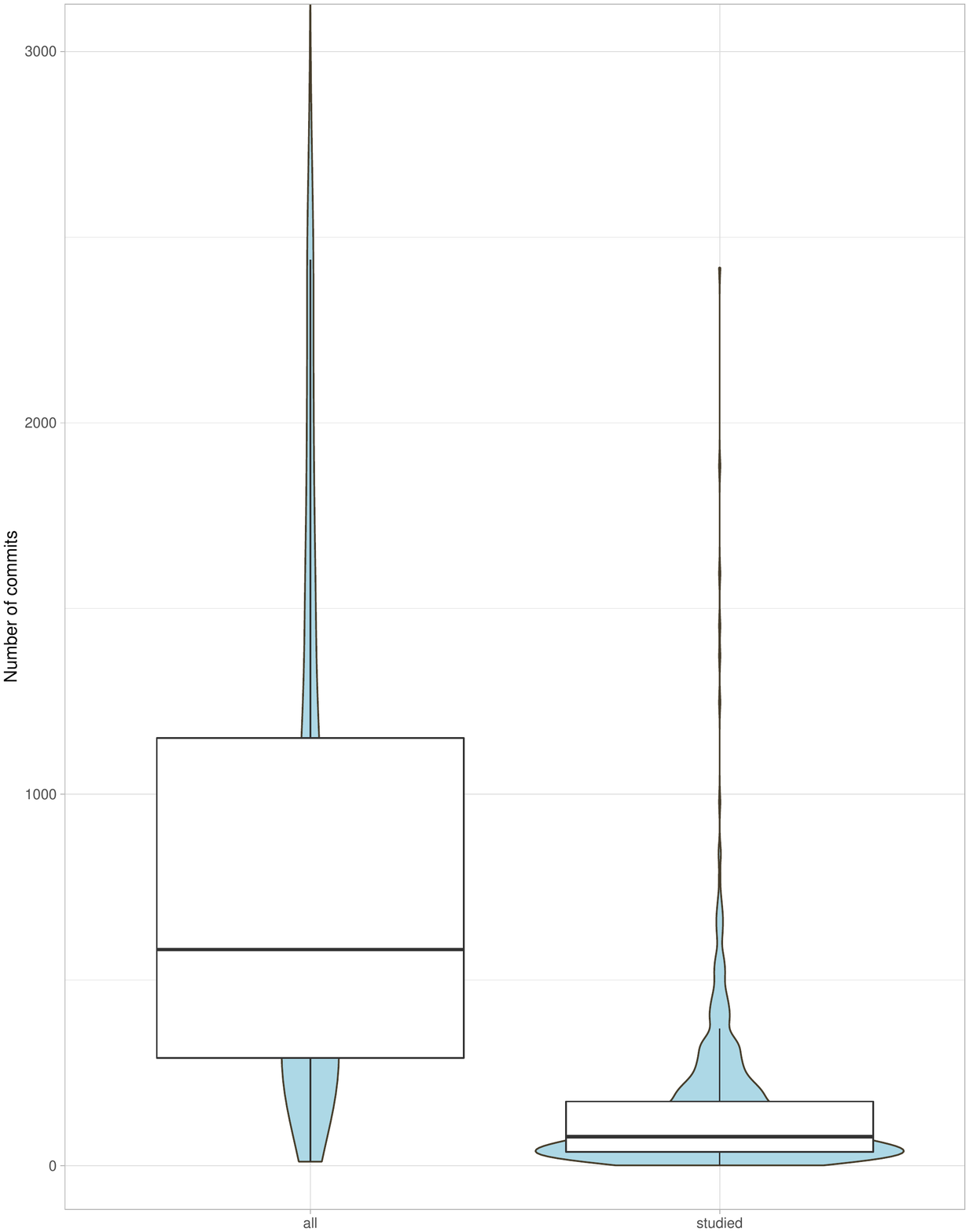}
 \label{Figure:commits}
 \caption{Commits}
\end{subfigure}
\begin{subfigure}{0.3\linewidth}
 \includegraphics[width=1.\linewidth]{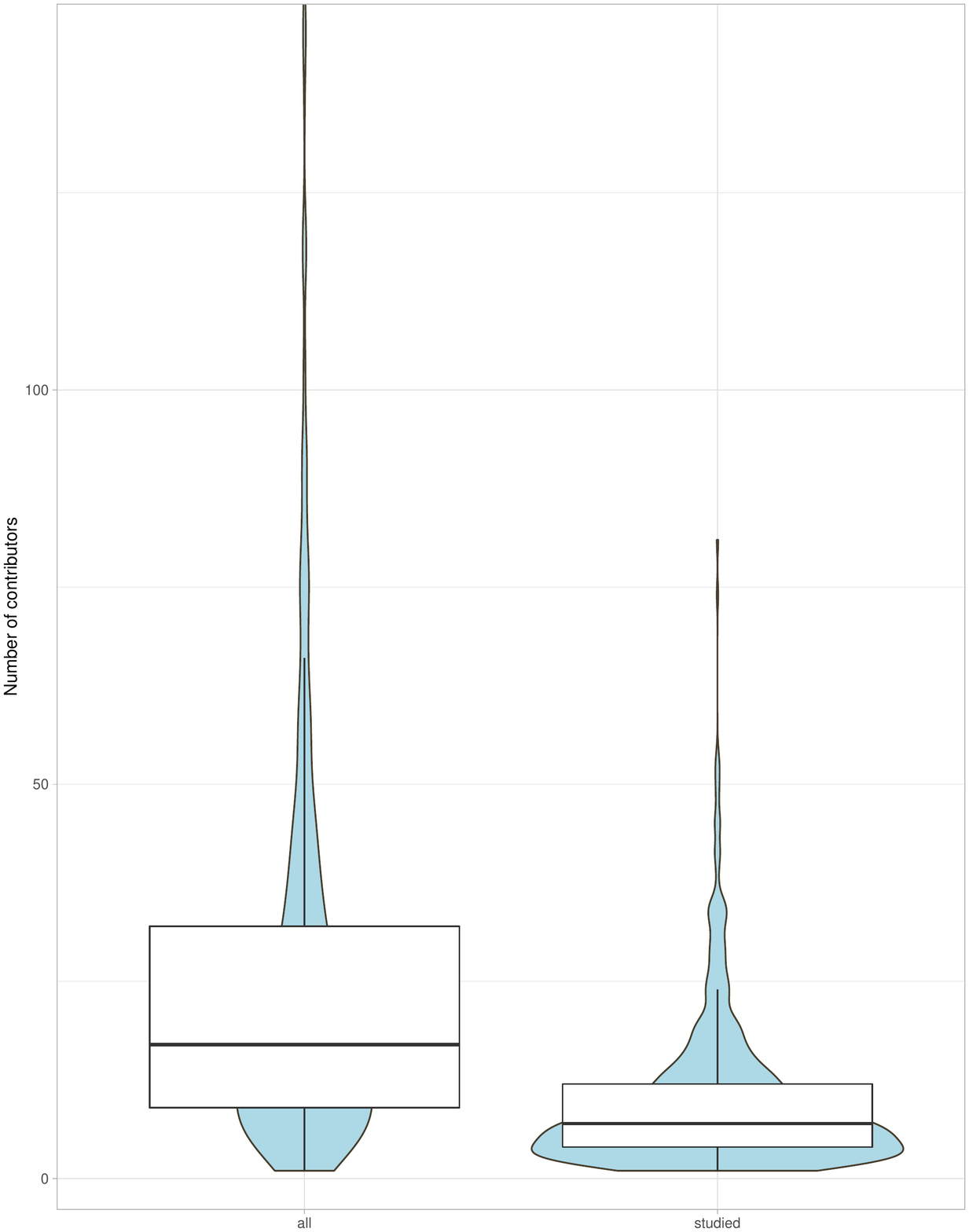} 
 \label{Figure:contributors}
 \caption{Contributors}
\end{subfigure}
\begin{subfigure}{0.3\linewidth}
 \includegraphics[width=1.\linewidth]{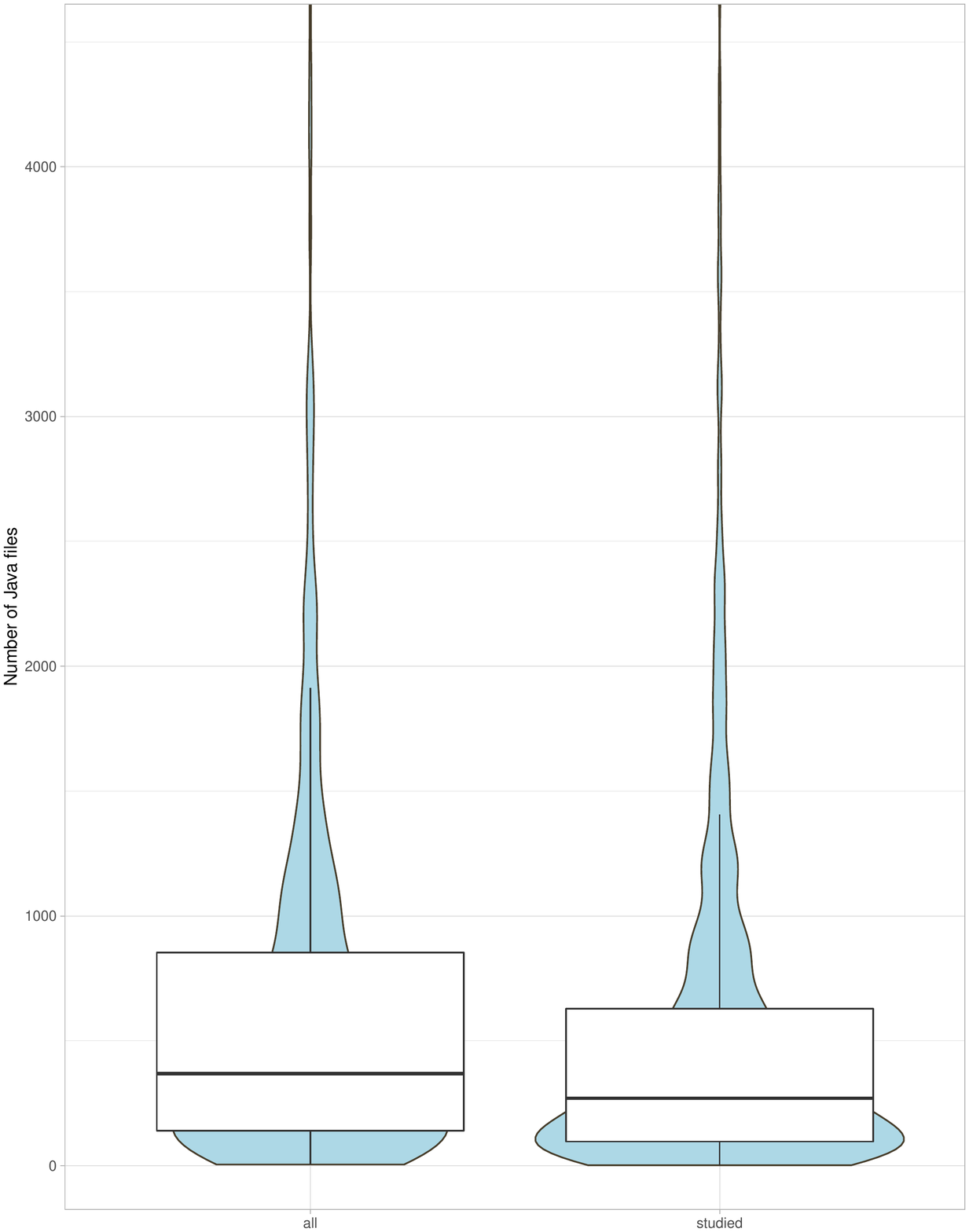} 
 \label{Figure:files}
 \caption{Source Files}
\end{subfigure}
\caption{Distribution of (a) commits, (b) contributors, and (c) Java source files of repositories}
\label{Figure:projectsc_violinplot}
\end{figure}

To conduct our exploratory study, we used a dataset of well-engineered open-source Java projects. The authors of this dataset \cite{munaiah2017curating} curated a set of open-source projects, proven to follow software engineering practices such as documentation, testing, issue and bug tracking, and project management. We chose this dataset as it was also analyzed in previous studies \cite{peruma2019context,alomar2019impact,fakhoury2019improving} that have been mining refactoring operations, just like our study. In total, our dataset is composed of 800 projects hosted on GitHub. Each project was cloned in order to extract the data needed for our experiments. This data included, but not limited to, each commit author, source files impacted by each committed change, and timestamps etc.  Figure \ref{Figure:projectsc_violinplot} shows violin plots with the distribution of commits, number of contributors, and size (number of .java files) of the selected repositories. We provide plots for all data of 800 systems (labeled as \textit{all}), and for a subset of data of 800 systems that effectively analyzed in the study (labeled as \textit{studied}), which correspond to the repositories with at least one refactoring detected in the commits during the study period. Table \ref{Table:Statistics} shows the statistics of our dataset. The projects in our dataset have 74.6\% of the projects (i.e., 597 projects) had their most recent commit within the last three years. Given that 597 out of the 800 monitored repositories were active during that period, we found refactoring activity in active projects is statistically significant than the ones in inactive projects with at least one refactoring commit detected during that period. An detailed overview of the studied project's is provided in Table~\ref{Table:DATA_Overview}.

\begin{table*}
\centering
\caption{Statistical summary of our dataset}
\label{Table:Statistics}
\begin{tabular}{@{}lrrrrrr|rrrrrr@{}}
\toprule
\multicolumn{1}{c}{\multirow{3}{*}{\textbf{Statistics}}} & \multicolumn{6}{c|}{\textit{\textbf{All}}} & \multicolumn{6}{c}{\textit{\textbf{Studied}}} \\ \cmidrule(l){2-13} 
\multicolumn{1}{c}{} & \multicolumn{1}{c}{\textbf{Min}} & \multicolumn{1}{c}{\textbf{Q1}} & \multicolumn{1}{c}{\textbf{Median}} & \multicolumn{1}{c}{\textbf{Mean}} & \multicolumn{1}{c}{\textbf{Q3}} & \multicolumn{1}{c|}{\textbf{Max}} & \multicolumn{1}{c}{\textbf{Min}} & \multicolumn{1}{c}{\textbf{Q1}} & \multicolumn{1}{c}{\textbf{Median}} & \multicolumn{1}{c}{\textbf{Mean}} & \multicolumn{1}{c}{\textbf{Q3}} & \multicolumn{1}{l}{\textbf{Max}} \\ \midrule
Commits & 11 & 290 & 582 & 935 & 1151.50 & 2439 & 1 & 37 & 78 & 139.85 & 173 & 370 \\
Contributors & 1 & 9 & 17 & 27.02 & 32 & 66 & 1 & 4 & 7 & 9.31 & 12 & 24 \\
Java files & 5 & 140.50 & 368.50 & 698.89 & 855 & 1913 & 3 & 96.50 & 270 & 528.26 & 628.50 & 1407\\
\midrule
\multicolumn{1}{c}{\multirow{3}{*}{\textbf{}}} & \multicolumn{6}{c|}{\textit{\textbf{Currently Active} (597 studied projects)}} & \multicolumn{6}{c}{\textit{\textbf{Currently Inactive} (203 studied projects)}} \\ \cmidrule(l){2-13} 
\multicolumn{1}{c}{} & \multicolumn{1}{c}{\textbf{Min}} & \multicolumn{1}{c}{\textbf{Q1}} & \multicolumn{1}{c}{\textbf{Median}} & \multicolumn{1}{c}{\textbf{Mean}} & \multicolumn{1}{c}{\textbf{Q3}} & \multicolumn{1}{c|}{\textbf{Max}} & \multicolumn{1}{c}{\textbf{Min}} & \multicolumn{1}{c}{\textbf{Q1}} & \multicolumn{1}{c}{\textbf{Median}} & \multicolumn{1}{c}{\textbf{Mean}} & \multicolumn{1}{c}{\textbf{Q3}} & \multicolumn{1}{l}{\textbf{Max}} \\ 
\midrule
Age (in days)  & 307 & 334 &  460 & 610.22 & 847 & 1459  & 1475 & 1743 & 1957 &2045  & 2325 & 3193 \\
\bottomrule
\end{tabular}
\end{table*}

\begin{table}[h]
\begin{center}
\caption{Projects overview}
\label{Table:DATA_Overview}
\begin{tabular}{lr}\hline
\toprule
\bfseries Item & \bfseries Count \\
\midrule
Total of projects & 800 \\
Total commits & 748,001 \\
Refactoring commits & 111,884 \\
Refactoring operations & 711,495 \\
\midrule 
\multicolumn{2}{c}{\textbf{\textit{Considered Projects - Refactored Code Elements}}}\\
\bfseries Code Element & \bfseries \# of Refactorings  \\
\midrule
Method & 222,785 \\
Attribute & 201,791  \\
Class & 121,625   \\
Variable & 115,717 \\
Parameter & 48,054\\
Package & 2380  \\
Interface & 1742 \\
\bottomrule
\end{tabular}
\end{center}
\end{table}

\subsection{Refactoring Detection}

Next, we utilized Refactoring Miner \cite{tsantalis2018accurate} to identify refactoring operations occurring in the projects. Refactoring Miner iterates over the commit history of a repository in chronological and compares the changes made to Java source code files in order to detect refactorings. Of the available state-of-the-art set of refactoring detection tools, Refactoring Miner has the highest performance, more specifically, a precision of 98\% and a recall of 87\% \cite{silva2016we,tsantalis2018accurate}.  To validate the detection of refactorings, we have also conducted a manual validation of the refactoring types identified by the Refactoring Miner tool. Such validation covered a random set of 30 refactoring commits, achieving a good performance. Running  Refactoring Miner on the projects under study, resulted scanning a total of 748,001 commits, from which, 111,884 commits contained at least one refactoring operation, and we collected a total of 711,495 refactoring operations. On average, each project contains 732 refactoring commits authored by 19 developers. 

\subsection{Commit Classification Model Construction} 
\label{subsec:commit classification}
\begin{table}
\centering
\caption{Classification categories}
\label{Table:ClassificationCategories}
\begin{tabular}{@{}ll@{}}
\toprule
\multicolumn{1}{c}{\textbf{Category}} & \multicolumn{1}{c}{\textbf{Description}} \\ \midrule
Functional & Feature implementation, modification or removal \\
Bug Fix & Tagging, debugging, and application of bug fixes \\
Internal QA & \begin{tabular}[c]{@{}l@{}}Restructuring and repackaging the system's code elements \\ to improve its internal design such as coupling and cohesion\end{tabular} \\
Code Smell Resolution & \begin{tabular}[c]{@{}l@{}}Removal of design defects that might violate the fundamentals \\ of software design principles and decrease code quality such \\ as duplicated code and long method\end{tabular} \\
External QA & \begin{tabular}[c]{@{}l@{}}Property or feature that indicates the effectiveness of a system \\ such as testability, understandability, and readability\end{tabular} \\ \bottomrule
\end{tabular}
\end{table}

After detecting all of the refactoring operations and the corresponding commits, the next step is to classify these commit messages into one of the refactoring motivations reported in \cite{alomar2020howwe}. Table \ref{Table:ClassificationCategories} shows five motivations that drive developers to refactor their code. The refactoring categories have been defined by reviewing the literature on refactoring motivation \citep{moser2006does,tsantalis2013multidimensional,silva2016we,palomba2017exploratory,alomar2019impact,pantiuchina2020developers,paixao2020behind,alomar2020reusability,alomar2020howwe,alomar2021xerox}. To cover all of the existing motivations, the authors clustered the existing refactoring taxonomy reported in the literature into five categories. We then followed a multi-staged approach to build our model for commit messages classification. The first stage consists of the model construction. In the second stage, we utilized the built  model to classify the entire dataset of commit messages. An overview of our methodology is depicted in Figure \ref{Figure:methodology}. In the following subsections, we detail the different steps in each stage.

\noindent \textbf{Model Construction.}
Our goal is to build a model from a corpus real world documented refactorings (i.e., commit message) to be utilized in the second stage to classify commit messages. The following subsections detail the different steps in the model construction phase.

\subsubsection{Data Annotation}
The model construction requires a gold set of labeled data to train and test the model. To prepare this set, a manual annotation of commit messages needs to be performed. To this end, we annotated 1,702 commit messages. This quantity roughly equates to a sample size with a confidence level of 95\% and a confidence interval of 2. The authors of this paper performed the annotation of the commit messages. Each author is provided with a random set of commit messages along with a detail definition of the annotation labels. Each annotator had to label each provided commit message with a label of either `Functional', `Bug Fix', `Internal Quality Attribute', `Code Smell Resolution', and `External Quality Attribute'. To mitigate bias in the annotation process, the annotated commit messages were peer-reviewed by the same group. All decisions made during the review had to be unanimous; discordant commit messages were discarded and replaced. In total, we annotated 348 commit messages as `Functional', `Bug Fix', `Internal Quality Attribute', and `Code Smell Resolution', while 310 messages were labeled as `External Quality Attribute'.

To avoid having false positive commits, we applied the filtering to narrow down the commit messages eliminating the ones that are less likely to be classified as one of the five motivation. We designed the filtering to help ensure that we only trained the algorithm on higher-quality commit messages \cite{jiang2017automatically}.

We followed the process from existing papers in filtering commit messages \cite{Mauczka2012,fu2015automated,da2017using}. For example, Fu et al. \cite{fu2015automated} filtered out short commit messages. Mauczka et al. \cite{Mauczka2012} used the \say{Blacklist} category to filter all commits, whose  underlying modifications were not carried out by humans or which do not actually include any source code modifications. In our work, we apply three filtering heuristics to narrow down the commit messages eliminating the ones that are less likely to be classified as one of the five categories. It is important to note that we removed short commit messages from the training, but not from the testing set because (1) short commit messages do not contain enough information and do not clearly describe the purpose of code change , and (2) we want to train the classifier on well-documented commit messages, and label commits that contain enough information about refactorings. Prior study has pruned short commit messages since these will be noise for the classifiers, and they did not record the cause of the changes \cite{fu2015automated}. Some criteria we used for filtering were as follows: 

\begin{itemize}
    \item Commits that were either too short or ambiguous were discarded. Some examples of hard-to-classify commit messages are: \say{\textit{Solr Indexer ready}}\footnote{https://github.com/01org/graphbuilder}, \say{\textit{allow multiple collections}}\footnote{https://github.com/0install/java-model}, and \say{\textit{Auto configuration of AgiScripts}}\footnote{https://github.com/1and1/attach-qar-maven-plugin}.
    \item If one commit could belong to more than one class, it was excluded.
    \item If the quality attribute is a part of the identifier name, the commits were excluded, e.g., \say{\textit{SONARJS-541 Precise issue location for ExpressionComplexity (S1067)}}.
    We discarded this commit because \say{complexity} is referring to a part of a class name and not a quality attribute.
    
\end{itemize}

The above-mentioned examples of ambiguous commit messages prevent us from being confident, and hence, for each discarded commit message, we randomly sampled another replacement.

\subsubsection{Text Pre-Processing}
We applied a similar methodology explained in \citep{kochhar2014automatic,le2015rclinker} for text pre-processing. In order for the commit messages to be classified into correct categories, they need to be preprocessed and cleaned; put into a format that the classification algorithms will accept. The activities involved in our pre-processing stage included: (1) expansion of word contractions (e.g., `I\textquotesingle m' $\rightarrow$ `I am'), (2) removal of URLs, single-character words, numbers, punctuation and non-alphabet characters, stop words, and (3) reducing each word to its lemma. The lemmatization process either replaces the suffix of a word with a different one or removes the suffix of a word to get the basic word form (lemma) \citep{lane2019natural}. In our work, the lemmatization process involves sentence separation, part-of-speech identification, and generating dictionary form. We split the commit messages into sentences, since input text could constitute a long chunk of text. The part-of-speech identification helps in filtering words used as features that aid in key-phrase extraction. Lastly, since the word could have multiple dictionary forms, only the most probable form is generated. We opted to use lemmatization over stemming, as the lemma of a word is a valid English word \citep{lane2019natural}. As for stopwords, we used the default set of stopwords supplied by NLTK \citep{Bird2002NLTKTN} and also added our own set of custom stop words. To derive the set of custom stop words, we generated and manually analyzed the set of frequently occurring words in our corpus.  Custom stop words include `git', `code', `refactor', `svn', etc. Additionally, for more effective pre-processing, we tokenized each commit message by splitting the text into its constituent set of words.

\subsubsection{Training and Test Data Preparation}
To gauge the accuracy of a machine learning model, the implemented model must be evaluated on a never-seen-before set of observations with known labels. Thus, the set of annotated commit messages were divided into two sub-datasets - a training set and a test set. The training set was utilized to construct the model while the test set was utilized to evaluate the classification ability of the model. For our experiment, we performed a shuffled stratified split of the annotated dataset. Our test dataset contained 25\% of the annotated commit messages, while the training dataset contained the remaining 75\% of annotated commit messages. This split results in the training dataset containing a total of 1,276 commit messages, which breaks down to 246 `Functional', 271 `BugFix', 255 `Internal', 276 `CodeSmell', and 228 `External' labeled commit messages. The stratification was performed based on the class of the commit messages. The use of a random stratified split ensures a better representation of the different types (i.e., labels) of commit messages and helps reduce the variability within the strata \citep{singh2013elements}.

\subsubsection{Feature Extraction}
After cleaning and preprocessing the commit message, we need to provide the classifier with a set of features that are associated with the commit messages in our dataset. However, not all features associated with each commit message will be useful in improving the prediction abilities of the model. Hence, a feature engineering task is required to determine the set of optimum features \citep{zheng2018feature}. In our study, we constructed our model using the text in the commit message. Hence, the feature for this model is limited to the commit message. We utilized Term Frequency-Inverse Document Frequency (TF-IDF) \citep{manning2008introduction}, commonly used in the literature \citep{lin2013empirical,le2015rclinker,alomar2020howwe,alomar2020toward},  to convert the textual data into a vector space model that can be passed into the classifier. In our experiments, we evaluate the accuracy of the model by constructing the TF-IDF vectors using different types of N-Grams and feature sizes. The N-Gram technique is a set of \textit{n-word} that occurs in a text set and could be used as a feature to represent that text \citep{kowsari2019text}. In our classification, we use N-Grams since it is very common to enhance the performance of text classification \citep{tan2002use}. Using TF-IDF, we can determine words that are common and rare across the documents (i.e., commit messages) in our dataset; the model utilizes these words. In other words, The value for each N-Gram is proportional to its TF score multiplied by its IDF score. Thus, each preprocessed word in the commit message is assigned a value which is the weight of the word computed using this weighting scheme.

\subsubsection{Model Training}

For our study, we evaluated the accuracy of six machine learning classifiers: Random Forest, Logistic Regression, Multinomial Naive Bayes, K-Nearest Neighbors, Support Vector Classification (C-Support Vector Classification based on LIBSVM \citep{cSupportVector,LIBSVM}), and Decision Tree (CART \citep{CART}). We selected these classifiers since they are widely adopted in several classification problems in software engineering (e.g.,\citep{Hindle:2011:ATN:1985441.1985466,kochhar2014automatic,Levin:2017:BAC:3127005.3127016,honel2019importance,alomar2020toward}).

It is important to note that the library containing the classification algorithms is capable of multiclass classification. As per the Python's SKlearn documentation, Random Forest, K-Nearest Neighbors, Logistic Regression, and Multinomial Naive Bayes are inherently multiclass \citep{inherently}, while SVC utilizes a one-vs-one approach to handle multiclass \citep{SVC}. Moreover, to ensure consistency, we ran each classifier with the same set of test and training data each time we updated the input features.

\subsubsection{Model Tuning \& Evaluation}
The goal of this step is to obtain the optimal set of classifier parameters that provide the highest performance by tuning the hyperparameters. For numeric-based hyperparameters, we determined the bounds/range for testing through continuously running the classifier with a different range of values to identify the appropriate minimum and maximum value. We performed our hyperparameter tuning on the training dataset using a combination of 10-fold cross-validation and an exhaustive grid search \citep{dangeti2017statistics}. Our test dataset did not take part in the training process, which provides a more realistic model evaluation. The combination of hyperparameters that resulted in the highest Micro-F1 score was selected to construct the model. Table \ref{Table:Parameters} provides the optimal hyperparameter values for the classification algorithms in our study.


\begin{table}[h!]
\centering
\caption{Optimal parameter values for the classification algorithms}
\label{Table:Parameters}
\begin{tabular}{@{}lll@{}}
\toprule
\multicolumn{1}{l}{\textbf{Algorithm}} & \multicolumn{1}{l}{\textbf{Parameter}} & \multicolumn{1}{l}{\textbf{Value}} \\ \midrule
\multirow{4}{*}{Random Forest} & max\_depth & 78 \\
 & n\_estimators & 500 \\
 & criterion & gini \\
 & bootstrap & false \\ \midrule
\multirow{3}{*}{Support Vector Classification} & c & 1.99 \\
 & gamma & scale \\
 & kernel & linear \\ \midrule
\multirow{2}{*}{Decision Tree} & criterion & gini \\
 & max\_depth & 75 \\ \midrule
\multirow{3}{*}{Logistic Regression} & penalty & l1 \\
 & solver & liblinear \\
 & c & 1.0 \\ \midrule
Multinomial Naive Bayes & alpha & 2.63 \\ \midrule
\multirow{2}{*}{K-Nearest Neighbors} & n\_neighbors & 69 \\
 & weights & uniform \\ \bottomrule
\end{tabular}
\end{table}

\subsubsection{Optimized Model}
In this stage, the optimized model produced by the training phase is utilized to predict the labels of the test dataset. Based on the predictions, we measure the precision and recall for each label as well as the overall F1-score of the model. In Section \ref{section:results}, we detail our classification results.

\noindent \textbf{Model Classification.}
We utilized the optimized model that we created in the prior stage. In order to be consistent, before classifying each commit message, we performed the same text pre-processing activities, as in the prior stage, on the commit message. The result of this stage is the classification of each refactoring commit into one of the five categories. The output of this classification process was utilized in our experiments in order to answer our corresponding research questions.

\subsection{Data Analysis}

Finally, we analyzed the output generated from our detection and extraction activities to answer our research questions. Since our research questions are both quantitative and qualitative, we used tools/scripts along with manual activities to arrive at our findings. For replication purposes, our dataset and other artifacts are available on our project website \cite{ProjectWebsite}.

\section{Results \& Discussion} \label{section:results}

In this section, we report and discuss our findings for analyzing the identified refactoring-related patterns to answer our research questions. For each research question, we defined the following hypothesis:

\begin{itemize}
    \item \textbf{Hypothesis \#1.} Whether developers with more contribution are most likely to have different motivations to refactor code.
    
    \textemdash \RQA
    
    \textemdash \RQB
    \item \textbf{Hypothesis \#2.} Whether developers with more contribution are most likely to be responsible for a higher number of refactoring activities.
    
    \textemdash \RQC
    \item \textbf{Hypothesis \#3.} Whether developers' experience correlates with better documentation of refactorings.
    
    \textemdash \RQD
\end{itemize}

\subsection{\RQA}

For our experiments on developer experience, we studied the project contributions made by the developer. In other words, we utilized the volume of commits made to source code files by a developer as a proxy for contribution. Introduced by~\cite{7972731},  this approach calculates the Developer’s Commit Ratio (DCR) for each developer in the project. This ratio measures the number of individual commits made by the developer against all project commits. It is worth noting that the DCR scores are normalized and it is comparable across projects as the values are not affected by the size of the projects. Formally, this ratio is defined as: 

 \begin{equation}
 DCR = \frac{\textit{Individual Contributor Commits}}{\textit{Total Project Commits}}
 \end{equation}

\begin{figure*}[]
 	\centering
 	\includegraphics[width=0.5\linewidth]{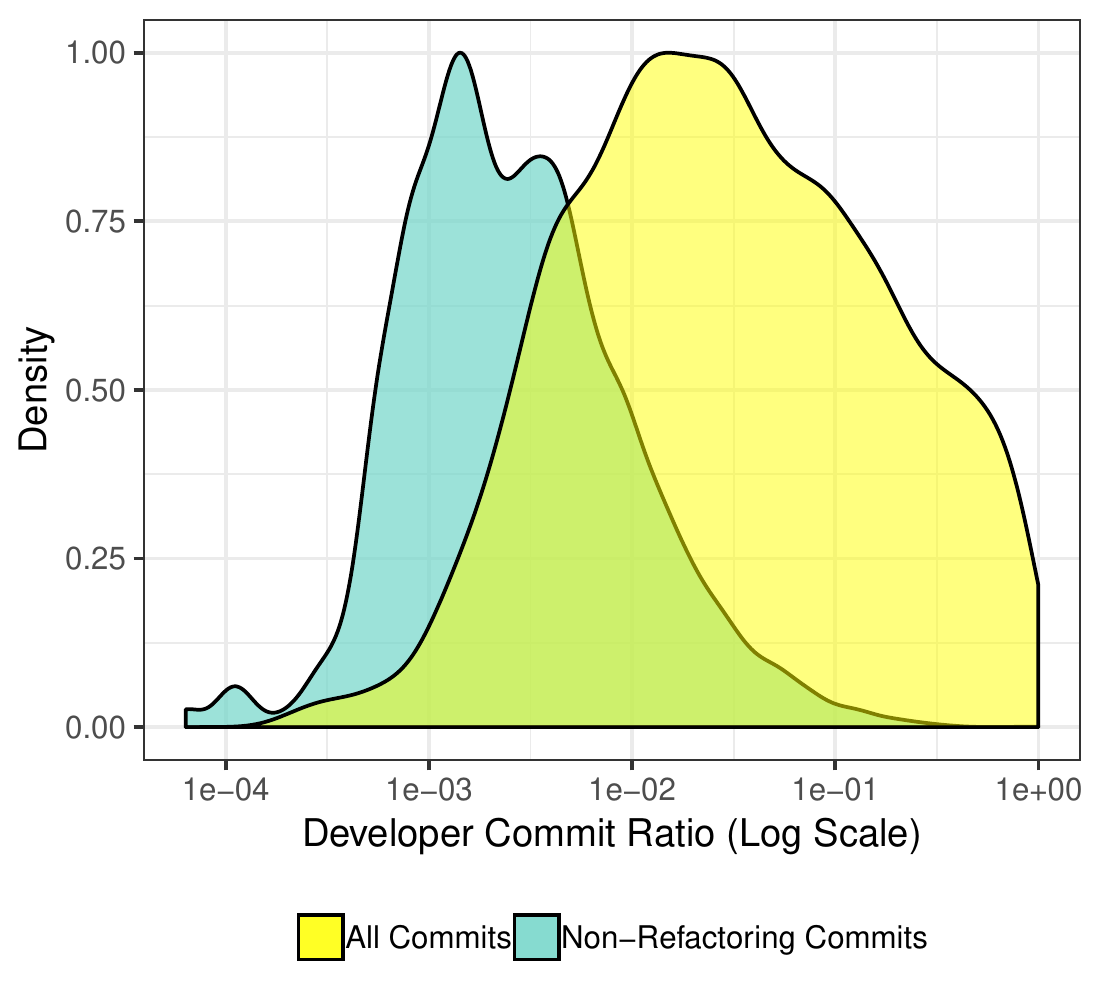}
 	\caption{Distribution of DCR values for developers based on the type of commit performed in their project}
 	\label{Figure:dcr}
\end{figure*}

\begin{table}[]
\centering
\caption{Statistical summary of DCR scores based on the type of commit performed in the project}
\label{Table:StatSummaryDCR}
\begin{tabular}{@{}crrrrr@{}}
\toprule
\textbf{Min} & \multicolumn{1}{c}{\textbf{Q1}} & \multicolumn{1}{c}{\textbf{Median}} & \multicolumn{1}{c}{\textbf{Mean}} & \multicolumn{1}{c}{\textbf{Q3}} & \multicolumn{1}{c}{\textbf{Max}} \\ \midrule
\multicolumn{6}{c}{\textit{Non-Refactoring Commits}} \\
\multicolumn{1}{r}{0.0001} & 0.0010 & 0.0019 & 0.0031 & 0.0043 & 0.0130 \\ \midrule
\multicolumn{6}{c}{\textit{All Commits}} \\
\multicolumn{1}{r}{0.0002} & 0.0065 & 0.0197 & 0.0456 & 0.0604 & 0.2632 \\ \bottomrule
\end{tabular}
\end{table}

In our experiment, we consider the author of a commit as its developer. We followed the same approach in Peruma et al.~\cite{peruma2019context} who studied the DCR distribution of developers that perform rename refactorings. As shown in Figure~\ref{Figure:dcr}, we analyzed two types of distributions (1) developers who only performed non-refactoring operations (depicted as `Non-Refactoring Commits' in the chart), and (2) developers who performed a mix of refactoring and non-refactoring operations on the source code (depicted as `All Commits' in the chart). Not surprisingly, our dataset had a large proportion of developers that performed a mix of refactoring and non-refactoring operations. These developers also had a higher DCR score.


We also observe from Figure~\ref{Figure:dcr} that developers who are tend to interleave refactorings in their commits, have a higher DCR score than developers whose commits do not contain refactorings. Even though we see an overlap in the density plot, the majority of non-refactoring developers are more concentrated on the lower end of the DCR scale. Furthermore, looking at the statistical summary of DCR scores in Table~\ref{Table:StatSummaryDCR}, we see that the average DCR score of a developer performing only non-refactoring commits is 0.0031 while the average DCR score of a developer performing all types of commit operations is 0.0456. Similar to~\cite{peruma2019context} we performed a non-parametric Mann-Whitney-Wilcoxon test on the DCR values for developers that do not perform any refactoring operations and those that did. We obtained a statistically significant p-value ($<0.05$) when the DCR values of these two groups of developers were compared.
Hence, this shows that developers working exclusively on new features to a system are more likely to have less contribution in the project than developers whose work also includes performing refactoring activities. Our findings also confirm the studies carried out by \cite{tsantalis2013multidimensional,kim2014empirical} that developer experience is an essential factor when it comes to software refactoring. However, unlike \cite{tsantalis2013multidimensional,kim2014empirical}, the approach we took relied on a metric (DCR) and was performed automatically over a much larger sample.

\vspace{.2cm}
\begin{tcolorbox}
\textit{Summary}.
Using an alternate approach (i.e., developer contributions), we confirm findings from prior research that developers with more contributions are typically involved in refactoring activities in systems. As our approach utilizes existing repository data, and is automated, it provides a non-subjective and scalable approach to estimate the most contributed developers in a project and thereby help to identify developers that are suitable for specific project tasks.
\end{tcolorbox}

\subsection{\RQB} 

In this research question, we investigate whether specific developers are significantly contributing to the overall refactoring of the system, or if it is randomly distributed among all developers. We approach this research question from two fronts -  quantitative and qualitative. In the quantitative approach, we perform an empirical and automated study on our dataset. In the qualitative approach, we perform a manual, case study like investigation on a select set of projects.

\begin{figure}[]
 \centering
\begin{subfigure}{0.48\linewidth}
 \includegraphics[width=1.\linewidth]{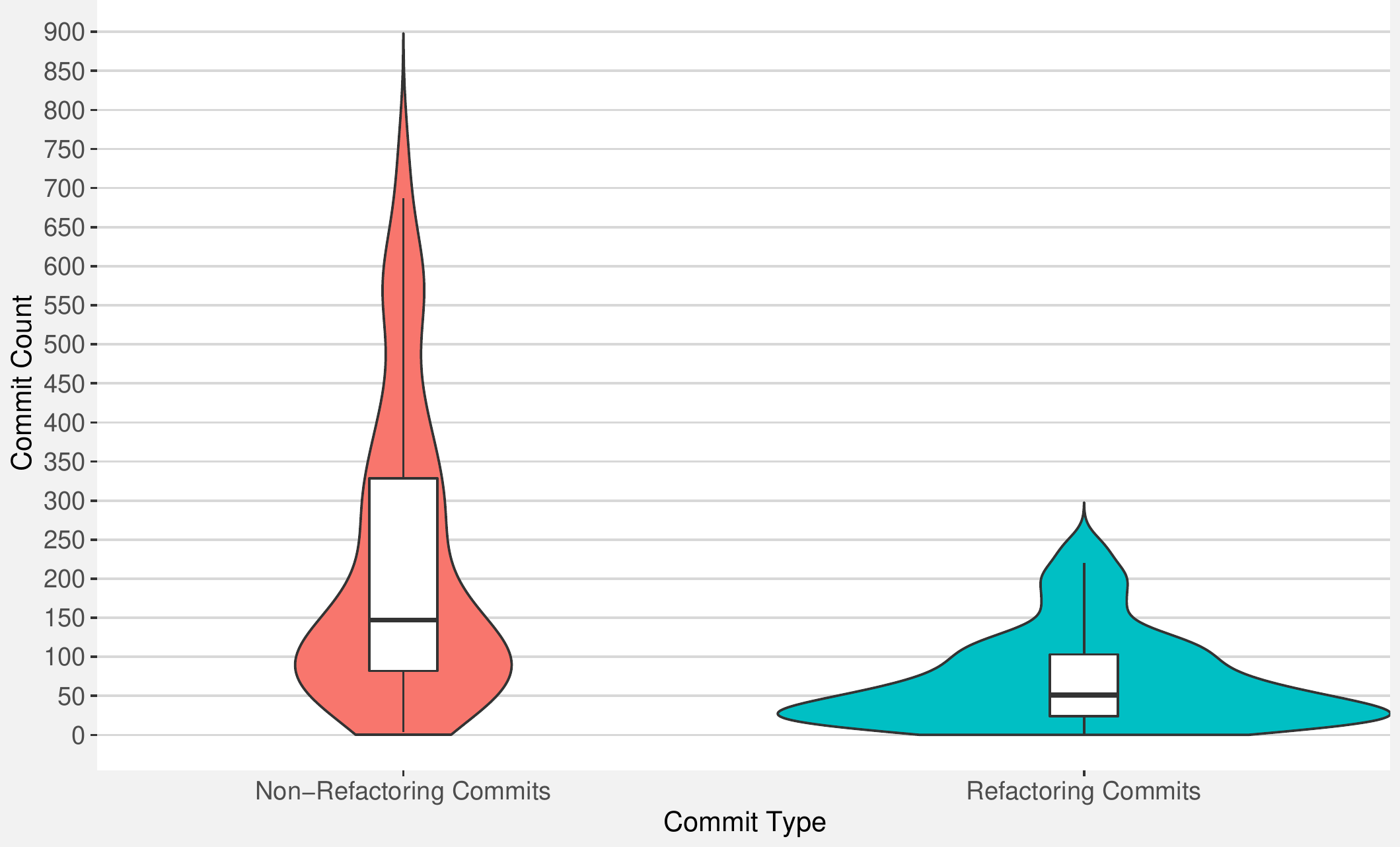}
 \label{Figure:violinA}
 \caption{Top 5\% of Developers}
\end{subfigure}
\begin{subfigure}{0.48\linewidth}
 \includegraphics[width=1.\linewidth]{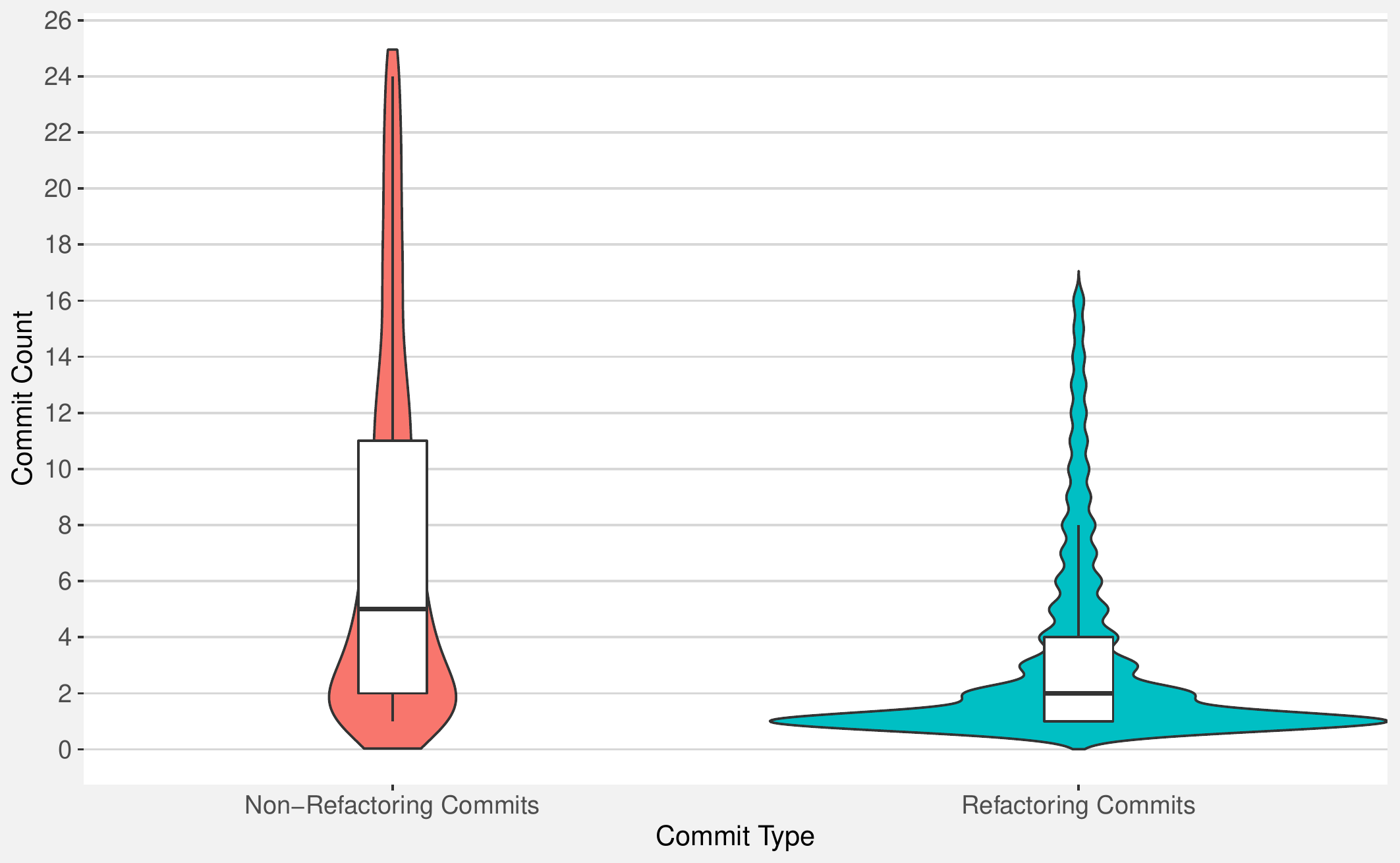} 
 \label{Figure:violinB}
 \caption{The Rest of Developers}
\end{subfigure}
\caption{Comparative counts of refactoring and non-refactoring commits for developers. Chart (a) is for the top 5\% of developers, while chart (b) is for the remaining developers.}
\label{Figure:violin}
\end{figure}

\vspace{3mm}
\noindent \textbf{\textit{Quantitative Analysis.}}
This part of the research question investigates the DCR values associated with each developer in the project, along with the total number of refactoring and non-refactoring commits made by the developer for only Java source files. To perform the comparison, we split the developers into two sets. The first set consisted of developers that fell into the top 5\% (labeled as \emph{TOP-5}) of DCR scores while the second set contained the remaining (i.e., 95\%) developers. The \emph{TOP-5} of developers equated to approximately a 95\% confidence level and confidence interval of 5. Represented by the \emph{TOP-5} are 372 developers, while the remaining developers amount to 7,066. For developers in each of the two sets, we obtained the count of refactoring and non-refactoring commits made by the developer. Figure~\ref{Figure:violin} shows a violin plot of this dataset.  Figure~\ref{Figure:violin}(a) shows the refactoring and non-refactoring commits of the \emph{TOP-5} of developers, while  Figure~\ref{Figure:violin}(b) shows the same counts for the remaining developers. A violin plot provides an ideal mechanism to represent our findings as they are useful in providing a visual comparison of multiple distributions. For better interpretation and visualization, we removed outliers from the data via the Tukey's fences approach \cite{tukey1977exploratory}.

Looking at Figure~\ref{Figure:violin}, the first observation is the volume of commit counts made by the two sets of developers. A majority of the \emph{TOP-5} developers contribute significantly more to the project in terms of refactoring and non-refactoring commits. On average, a \emph{TOP-5} developer makes 70.24 and 223.7 refactoring and non-refactoring commits, respectively. On the other hand, the rest of the developers average around 3.21 and 15.69 refactoring and non-refactoring commits, respectively. Furthermore, the \emph{TOP-5} violin plot shows a high frequency of developers performing, approximately, 15 to 75 refactoring commits. The same does not hold for non-refactoring commits, where we see a higher density within the range of 75 to 125 commits. Additionally, we observed that our dataset contains some developers that perform at most around 300 refactoring commits while non-refactoring commits go up to around 800. Hence, non-refactoring commit counts have a higher variation than refactoring commits. The refactoring box plot is more condensed than the non-refactoring boxplot; this indicates that the data varies less and hence is more consistent.

\begin{table}[]
\centering
\caption{Statistical summary of the volume of refactoring operations performed by the top 5\% and the remaining set of developers}
\label{Table:StatSummary}
\begin{tabular}{@{}crrrrr@{}}
\toprule
\textbf{Min} & \multicolumn{1}{c}{\textbf{Q1}} & \multicolumn{1}{c}{\textbf{Median}} & \multicolumn{1}{c}{\textbf{Mean}} & \multicolumn{1}{c}{\textbf{Q3}} & \multicolumn{1}{c}{\textbf{Max}} \\ \midrule
\multicolumn{6}{c}{\textit{Top 5\%}} \\
\multicolumn{1}{r}{1.00} & 2.00 & 6.00 & 11.14 & 15.00 & 55.00 \\ \midrule
\multicolumn{6}{c}{\textit{Rest}} \\
\multicolumn{1}{r}{1.00} & 1.00 & 2.00 & 3.57 & 5.00 & 16.00 \\ \bottomrule
\end{tabular}
\end{table}

\begin{figure}[]
 	\centering
 	\includegraphics[width=0.5\textwidth]{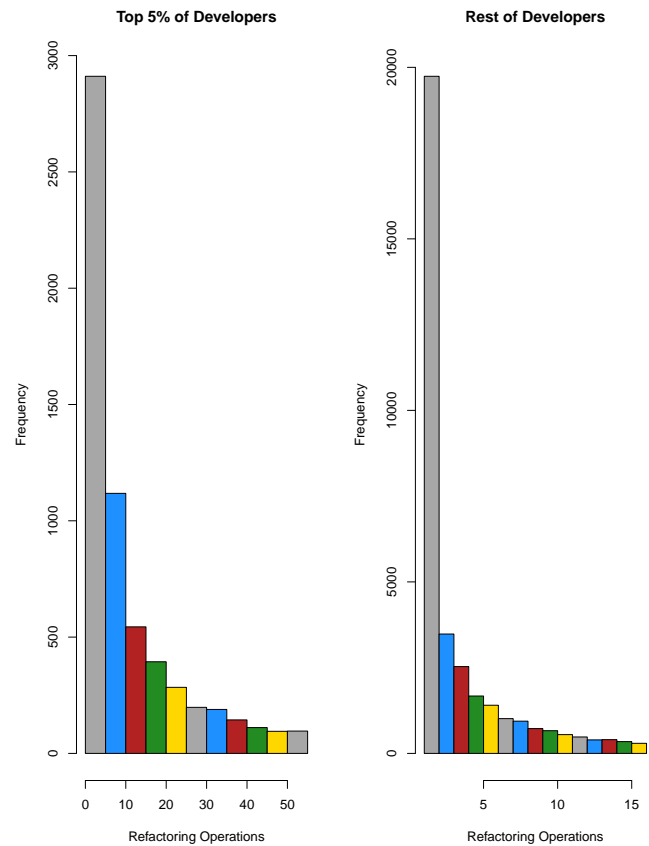}
 	\caption{Histogram of refactoring operations performed by the top 5\% and the remaining set of developers}
 	\label{Figure:histogram}
\end{figure}

Finally, we looked at the number of refactoring operations performed by the two groups of developers. It should be noted that a single refactoring commit can contain one or more refactoring operations. A statistical summary of our findings is presented in Table~\ref{Table:StatSummary}, while a comparative histogram is available in Figure~\ref{Figure:histogram}. Even though the histogram shows a higher volume of refactoring operations by less contributed developers, it should be noted that this is the cumulative count across all projects in the dataset. If we were to look at the individual developer contributions, we could see that more contributed developers apply refactorings more often than less contributed developers.

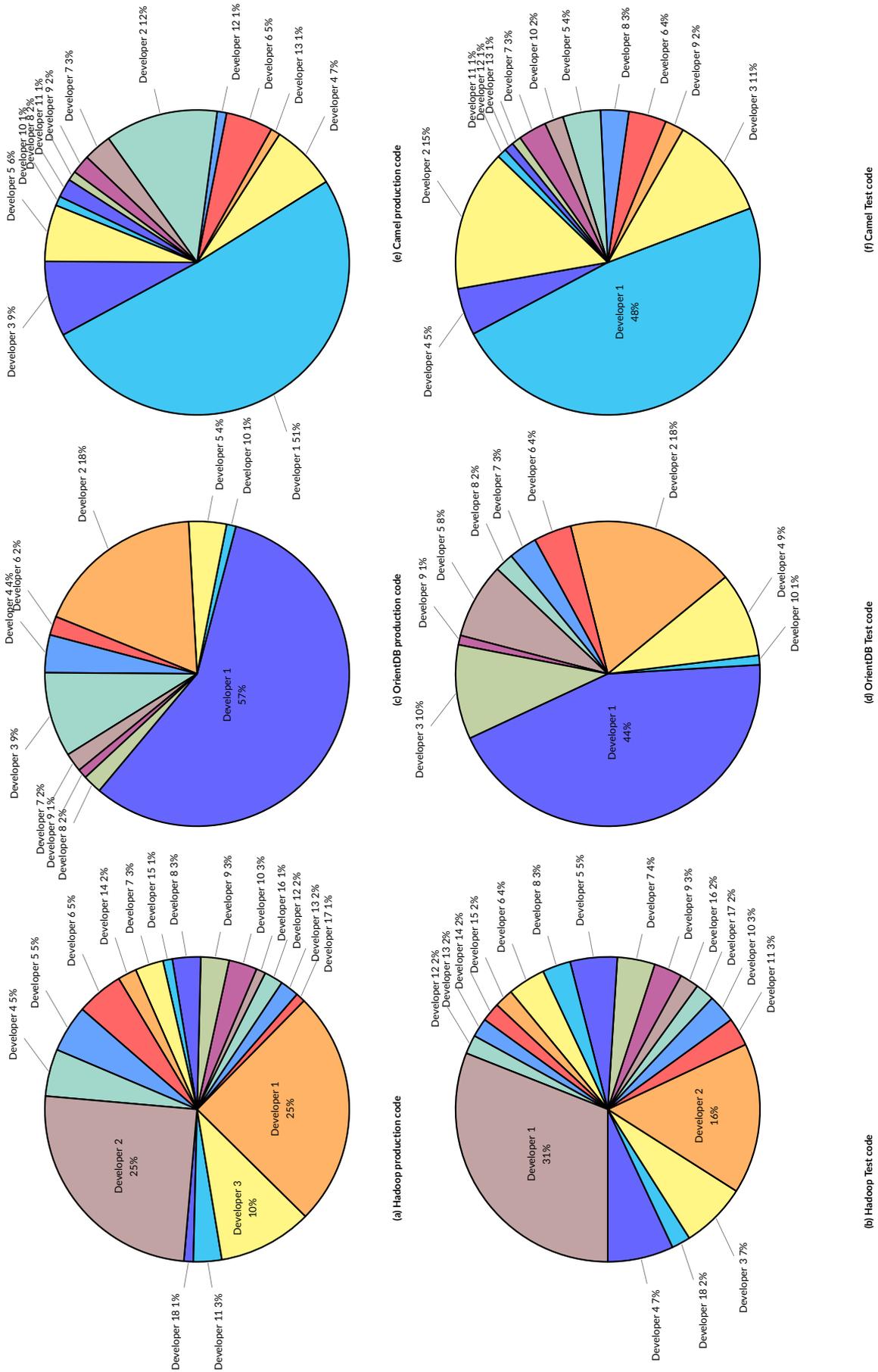
\begin{figure*}[tbp]
\centering 
\caption{Refactoring contributors in production and test files in Hadoop, OrientDB and Camel}
\label{fig:Refactoring_Contributors}
\begin{sideways}
\begin{tikzpicture}[font=\tiny]
\begin{scope}[scale=0.42]

\pie[rotate = 175,radius =6.3,text=inside,outside under=10,no number]{1/Developer 18\and1\%, 3/Developer 11\and3\%, 10/Developer 3\and10\%, 25/Developer 1\and25\%, 1/Developer 17\and1\%, 2/Developer 13\and2\%, 2/Developer 12\and2\%, 1/Developer 16\and1\%, 3/Developer 10\and3\%, 3/Developer 9\and3\%, 3/Developer 8\and3\%,1/Developer 15\and1\%,3/Developer 7\and3\%,2/Developer 14\and2\%,5/Developer 6\and5\%,5/Developer 5\and5\%,5/Developer 4\and5\%,25/Developer 2\and25\%}{Hadoop Production Code}


\pie[rotate = 140,radius =6.3,pos ={18,0},text=inside,outside under=17,no number]{57/Developer 1\and57\%, 1/Developer 10\and1\%, 4/Developer 5\and4\%, 18/Developer 2\and18\%, 2/Developer 6\and2\%,4/Developer 4\and4\%, 9/Developer 3\and9\%, 2/Developer 7\and2\%, 1/Developer 9\and1\%, 2/Developer 8\and2\%}{OrientDB Production Code}


\pie[rotate = 86,radius =6.3,pos ={35,0},text=inside,outside under=10,no number]{9/Developer 3\and9\%, 51/Developer 1\and51\%, 7/Developer 4\and7\%, 1/Developer 13\and1\%, 5/Developer 6\and5\%, 1/Developer 12\and1\%, 12/Developer 2\and12\%, 3/Developer 7\and3\%, 2/Developer 9\and2\%, 1/Developer 11\and1\%, 2/Developer 8\and2\%, 1/Developer 10\and1\%, 6/Developer 5\and6\%}


\pie[rotate = 180,radius =6.3,pos ={0,-17},text=inside,outside under=30,no number]{7/Developer 4\and7\%, 2/Developer 18\and2\%, 7/Developer 3\and7\%, 16/Developer 2\and16\%, 3/Developer 11\and3\%, 3/Developer 10\and3\%, 2/Developer 17\and2\%, 2/Developer 16\and2\%, 3/Developer 9\and3\%, 4/Developer 7\and4\%, 5/Developer 5\and5\%,3/Developer 8\and3\%,4/Developer 6\and4\%,2/Developer 15\and2\%,2/Developer 14\and2\%,2/Developer 13\and2\%,2/Developer 12\and2\%,31/Developer 1\and31\%}{Hadoop Test Code}


\pie[rotate = 115,radius =6.3,pos ={18,-17},text=inside,outside under=17,no number]{44/Developer 1\and44\%, 1/Developer 10\and1\%, 9/Developer 4\and9\%, 18/Developer 2\and18\%, 4/Developer 6\and4\%,3/Developer 7\and3\%, 2/Developer 8\and2\%, 8/Developer 5\and8\%, 1/Developer 9\and1\%,10/Developer 3\and10\%}{OrientDB Test Code}


\pie[rotate = 100,radius =6.3,pos ={35,-17},text=inside,outside under=10,no number]{5/Developer 4\and5\%, 48/Developer 1\and48\%, 11/Developer 3\and11\%, 2/Developer 9\and2\%, 4/Developer 6\and4\%, 3/Developer 8\and3\%, 4/Developer 5\and4\%, 2/Developer 10\and2\%, 3/Developer 7\and3\%, 1/Developer 13\and1\%, 1/Developer 12\and1\%, 1/Developer 11\and1\%, 15/Developer 2\and15\%}{Camel Test Code}

\centering
\node[below=33mm] {\hspace{420pt} \textbf{(a) Hadoop production code} \hspace{190pt} \textbf{(c) OrientDB production code} \hspace{150pt} \ \textbf{(e) Camel production code}};

\centering
\node[below=115mm] {\hspace{405pt} \textbf{(b) Hadoop Test code} \hspace{210pt} \textbf{(d) OrientDB Test code} \hspace{170pt}  \textbf{(f) Camel Test code}};

\end{scope}
\end{tikzpicture}
\end{sideways}

\end{figure*}

\vspace{4mm}
\noindent \textbf{\textit{Qualitative Analysis.}}
To better understand the key role of the \emph{TOP-5} contributors in the development team, we extract refactorings from a select set of projects - Hadoop\footnote{\url{https://github.com/apache/hadoop}}, OrientDB\footnote{\url{https://github.com/orientechnologies/orientdb}}, and Camel\footnote{\url{https://github.com/apache/camel}}. These three systems were randomly selected based on the criteria used in \cite{Levin:2017:BAC:3127005.3127016} (i.e., had more than 100 stars, had more than 60 forks, had size over 2 MB, these repositories are active and well-used). 
Next, we cluster production and test files of these projects, by developer ID. Finally, we carefully examine the top contributor's professional profiles to identify their role in the organization hosting the software project. Our findings are detailed below.

Figure~\ref{fig:Refactoring_Contributors}  
portrays the distribution of the refactoring activities on production code and test code performed by project contributors for each software system we examined. The Hadoop project has a total of 114 developers. Among them are 73 (64\%) refactoring contributors. As we observe in Figures~\ref{fig:Refactoring_Contributors}a 
and~\ref{fig:Refactoring_Contributors}b  
, not all of the developers are major refactoring contributors. The main refactoring contributor has a refactoring ratio of 25\% on production code and 10\% on test code. 
Figure~\ref{fig:Refactoring_Contributors}c  
and~\ref{fig:Refactoring_Contributors}d  
present the percentage of the refactorings for the OrientDB production code and test code. Out of the total 113 developers, 35 (31\%) were involved refactoring. The top contributor has a refactoring ratio of 57\% and 44\% on production and test code respectively.
For Camel, in Figures~\ref{fig:Refactoring_Contributors}e and~\ref{fig:Refactoring_Contributors}f 
, 73 (20\%) developers were on the refactoring list out of 368 total committers. The most active refactoring contributor has high ratios of 51\% and 48\% respectively in production and test code. We also note that very few developers applied refactorings exclusively on either production code or test code for the three projects under study.

The manual analysis aligns with the findings of the previous section in distinguishing a subset of developers that monopolize the refactoring activity across the three projects. To identify their key role in the development of the project, we searched, using their GitHub IDs, their professional profiles on Linked-In\footnote{Used in previous studies as a source to identify developers skills and experience.}. We were successful in locating the role of the top contributors for the 3 projects, and we found, through their public affiliation to the project, that they were either development leads or senior developers.

Our findings show that refactoring activities, in the 3 projects, are mainly performed by a subset 
of developers who have a management role in the company. 
Senior developers care more about refactoring the source code to ensure high-quality software and make the software easier for future development. These subsets of developers may perform certain practices when applying code refactoring (e.g., refactoring before and after adding new features, testing frequently to avoid any bugs that may introduce and affect the functionality of the software, and documenting and automating the application of refactoring). One of the reasons that seems not to encourage the other subsets of developers to significantly refactor the code is the technical constraints such as inadequate tool supports or lack of trust of automated support for composite refactorings. A discussion about various barriers to refactoring has been highlighted in Murphy-Hill et al. \cite{murphy2008breaking}. 

\vspace{.2cm}
\begin{tcolorbox}
\textit{Summary}. While refactorings are applied by various developers, only a reduced set of developers are responsible for performing the majority of these activities, in both production and test files. This set of developers take over refactoring activities without necessarily being dominant in other programming activities. As we examine the top contributor's publicly accessible professional profiles, we identify their positions to be advanced in the development team; hence, demonstrating their extensive knowledge of the design of the systems they contribute to.
\end{tcolorbox}

\subsection{\RQC}

\begin{table}

\centering
\caption{Detailed classification metrics (Precision, Recall, and F-measure) of each classifier}
\label{Table:classificationmetrics}
\centering
\begin{tabular}{lrrrlrrrlrrr}
\hline
\multicolumn{4}{|c|}{\textit{\textbf{Random Forest}}} & \multicolumn{4}{c|}{\textit{\textbf{Support Vector Classification}}} & \multicolumn{4}{c|}{\textit{\textbf{Decision Tree}}} \\ \hline
\multicolumn{1}{c}{\textbf{Category}} & \multicolumn{1}{c}{\textbf{Precision}} & \multicolumn{1}{c}{\textbf{Recall}} & \multicolumn{1}{c|}{\textbf{F1}} & \multicolumn{1}{c}{\textbf{Category}} & \multicolumn{1}{c}{\textbf{Precision}} & \multicolumn{1}{c}{\textbf{Recall}} & \multicolumn{1}{c|}{\textbf{F1}} & \textbf{Category} & \multicolumn{1}{l}{\textbf{Precision}} & \multicolumn{1}{l}{\textbf{Recall}} & \multicolumn{1}{l}{\textbf{F1}} \\ \hline
Bug Fix & 0.83 & 0.79 & \multicolumn{1}{r|}{0.81} & Bug Fix & 0.75 & 0.78 & \multicolumn{1}{r|}{0.77} & Bug Fix & 0.77 & 0.80 & 0.78 \\
Code Smell & 0.93 & 0.95 & \multicolumn{1}{r|}{0.94} & Code Smell & 0.93 & 0.94 & \multicolumn{1}{r|}{0.93} & Code Smell & 0.89 & 0.91 & 0.90 \\
External QA & 0.85 & 0.91 & \multicolumn{1}{r|}{0.88} & External QA & 0.92 & 0.89 & \multicolumn{1}{r|}{0.90} & External QA & 0.77 & 0.90 & 0.83 \\
Functional & 0.81 & 0.91 & \multicolumn{1}{r|}{0.86} & Functional & 0.77 & 0.88 & \multicolumn{1}{r|}{0.82} & Functional & 0.92 & 0.83 & 0.87 \\
Internal QA & 0.95 & 0.81 & \multicolumn{1}{r|}{0.87} & Internal QA & 0.95 & 0.84 & \multicolumn{1}{r|}{0.89} & Internal QA & 0.91 & 0.80 & 0.85 \\ \hline
Average F1 & 0.87 & 0.87 & \multicolumn{1}{r|}{0.87} & Average F1 & 0.87 & 0.86 & \multicolumn{1}{r|}{0.86} & Average F1 & 0.85 & 0.85 & 0.85 \\ \hline
\multicolumn{12}{l}{} \\ \hline
\multicolumn{4}{|c|}{\textit{\textbf{Logistic Regression}}} & \multicolumn{4}{c|}{\textit{\textbf{Multinomial Naive Bayes}}} & \multicolumn{4}{c|}{\textit{\textbf{K-Nearest Neighbors}}} \\ \hline
\multicolumn{1}{c}{\textbf{Category}} & \multicolumn{1}{c}{\textbf{Precision}} & \multicolumn{1}{c}{\textbf{Recall}} & \multicolumn{1}{c|}{\textbf{F1}} & \multicolumn{1}{c}{\textbf{Category}} & \multicolumn{1}{c}{\textbf{Precision}} & \multicolumn{1}{c}{\textbf{Recall}} & \multicolumn{1}{c|}{\textbf{F1}} & \multicolumn{1}{c}{\textbf{Category}} & \multicolumn{1}{c}{\textbf{Precision}} & \multicolumn{1}{c}{\textbf{Recall}} & \multicolumn{1}{c}{\textbf{F1}} \\ \hline
Bug Fix & 0.66 & 0.70 & \multicolumn{1}{r|}{0.68} & Bug Fix & 0.63 & 0.77 & \multicolumn{1}{r|}{0.69} & Bug Fix & 0.62 & 0.71 & 0.66 \\
Code Smell & 0.89 & 0.94 & \multicolumn{1}{r|}{0.91} & Code Smell & 0.82 & 0.94 & \multicolumn{1}{r|}{0.87} & Code Smell & 0.76 & 0.93 & 0.84 \\
External QA & 0.88 & 0.88 & \multicolumn{1}{r|}{0.88} & External QA & 0.97 & 0.71 & \multicolumn{1}{r|}{0.82} & External QA & 0.85 & 0.75 & 0.79 \\
Functional & 0.77 & 0.87 & \multicolumn{1}{r|}{0.82} & Functional & 0.66 & 0.83 & \multicolumn{1}{r|}{0.74} & Functional & 0.68 & 0.73 & 0.71 \\
Internal QA & 0.96 & 0.78 & \multicolumn{1}{r|}{0.86} & Internal QA & 0.99 & 0.67 & \multicolumn{1}{r|}{0.80} & Internal QA & 0.97 & 0.71 & 0.82 \\ \hline
Average F1 & 0.83 & 0.83 & \multicolumn{1}{r|}{0.83} & Average F1 & 0.81 & 0.78 & \multicolumn{1}{r|}{0.78} & Average F1 & 0.78 & 0.77 & 0.76 \\ \hline
\end{tabular}
\end{table}

To answer this research question, we present the refactoring commit messages classification results explained in Subsection~\ref{subsec:commit classification}. This section details the classification of 111,884 commit messages containing 711,495 refactoring operations. 
The complete set of scores for all the classifiers including the Precision, Recall, and F-measure scores per class for each machine learning classifier is provided in Table \ref{Table:classificationmetrics}. The best performing model was used to classify the test dataset. Based on our findings, we observed that \textbf{\textit{Random Forest achieved the best F1 score: 87\%}} which is higher than its competitors. Random Forest belongs to the family of ensemble learning machines, and has typically yielded superior predictive performance mainly due to the fact that it aggregates several learners. Hence, we utilized this machine learning algorithm (and its optimal set of hyperparameters) as the optimum model for our study.

\begin{figure*}[h]  
\centering 
\begin{tikzpicture}
\begin{scope}[scale=0.9]
\begin{axis}[
    ybar stacked,
    height=11cm,
    width=18cm,
    percentage plot,
    bar width=0.55cm, 
    xticklabels from table={\dataA}{Category}, 
    xtick=data,
    x tick label style={
      rotate=45,
      anchor=east, 
      xshift=-1.5mm, yshift=-2mm
    },
    legend style={
      at={(0.5,-0.44)},
      anchor=south,
      legend columns=-1
      },
]

    \addplot [fill=red!60]  table[percentage series=1] {\dataA};
    \addplot [fill=orange!60]                 table[percentage series=2] {\dataA};
    \addplot [fill=yellow!60]       table[percentage series=3] {\dataA};
    \addplot [fill=cyan!60]       table[percentage series=4] {\dataA};
    \addplot [fill=blue!60]       table[percentage series=5] {\dataA};


    \legend{\strut BugFix, \strut Functional, Internal QA, External QA, Code Smell}
\end{axis}
\end{scope}
\end{tikzpicture}
\caption{Distribution of refactoring types per category performed by top 5\% developers}
\label{fig:ref_dist_class_top5developers}
\end{figure*}
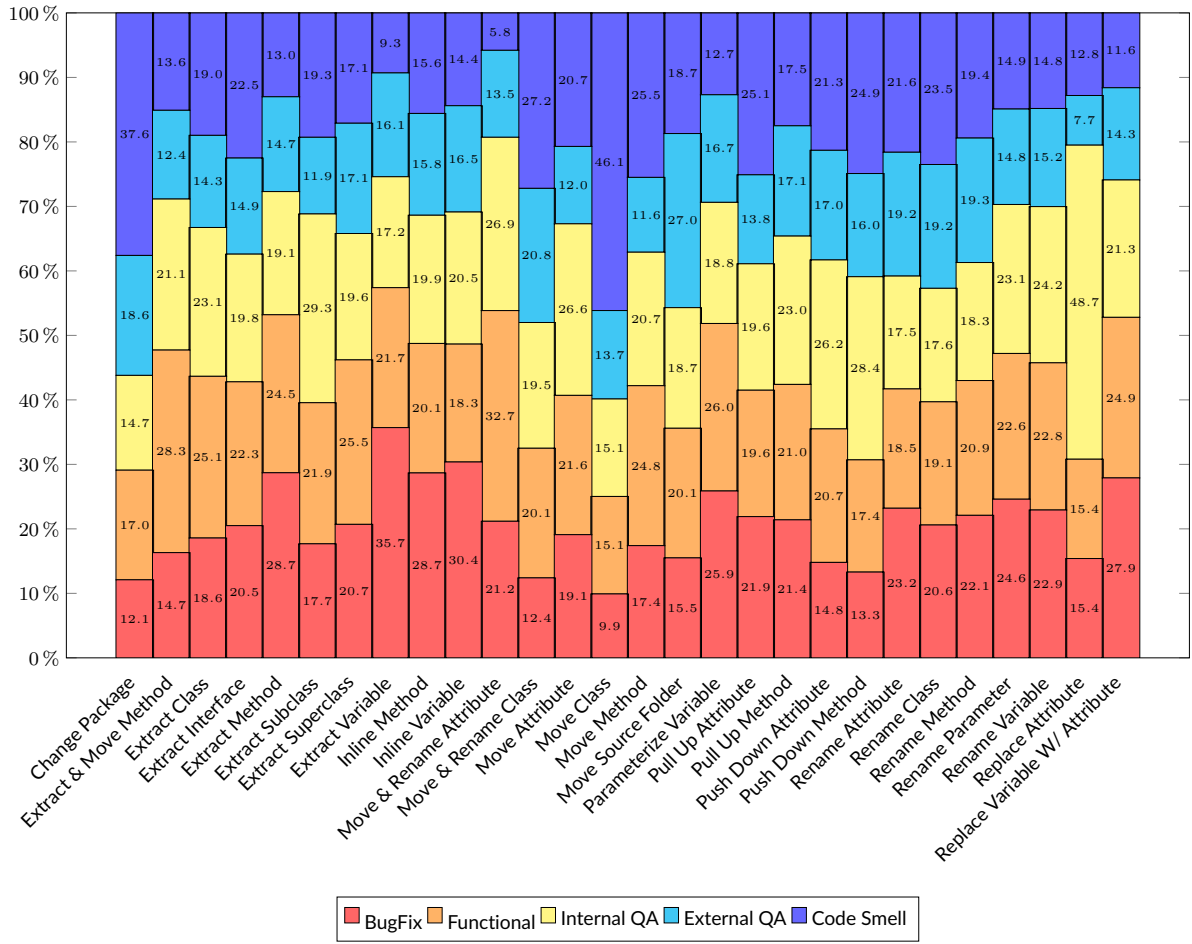

\pgfplotsset{compat=1.14}
\definecolor{findOptimalPartition}{HTML}{696969}
\definecolor{storeClusterComponent}{HTML}{808080}
\definecolor{dbscan}{HTML}{BEBEBE}
\definecolor{constructCluster}{HTML}{DCDCDC}

\pgfplotstableread[col sep=comma,header=true]{
Motivation,1,2,3,4,5
Change Package, 0, 0, 0, 100, 0
Extract \& Move Method, 11.1, 20.0, 35.6, 17.8, 15.6
Extract Class, 25.9, 37.0, 7.4, 11.1, 18.5
Extract Interface, 0, 0, 0, 0, 1
Extract Method, 14.8, 43.8, 8.6, 11.7, 21.1
Extract Subclass, 0, 50.0, 0, 0, 50.0
Extract Superclass, 11.8, 35.3, 41.2, 5.9, 5.9
Extract Variable, 39.1, 26.6, 4.7, 10.9, 18.8
Inline Method, 28.4, 22.6, 9.7, 6.5, 12.9
Inline Variable, 60.0, 20.0, 0, 20.0, 0
Move \& Rename Attribute, 0, 0, 0, 0, 0
Move \& Rename Class, 23.9, 40.8, 8.5, 0, 26.8
Move Attribute, 6.6, 19.8, 48.1, 11.3, 14.2
Move Class, 6.4, 9.2, 8.3, 6.4, 69.7
Move Method, 26.3, 26.3, 13.7, 4.2, 29.5
Move Source Folder, 50.0, 0, 0, 50.0, 0
Parameterize Variable, 60.0, 10.0, 0, 20.0, 10.0
Pull Up Attribute, 21.4, 64.3, 0, 14.3, 0
Pull Up Method, 12.2, 20.0, 67.8,0, 0
Push Down Attribute, 0, 3, 0, 0, 0
Push Down Method, 0, 50.0, 12.5, 12.5, 25.0
Rename Attribute, 14.7, 0, 0, 16.7, 68.4
Rename Class, 8.6, 17.3, 18.5, 19.8, 35.8
Rename Method, 23.0, 20.8, 22.3, 12.5, 21.4
Rename Parameter, 56.3, 12.3, 19.8, 3.2, 8.3
Rename Variable, 17.8, 33.2, 0.9, 4.9, 43.2
Replace Attribute, 0, 0, 0, 0, 0
Replace Variable W/ Attribute, 0, 0, 11, 0, 0
}\dataA

\pgfplotstablecreatecol[
 create col/expr={
    \thisrow{1} + \thisrow{2} + \thisrow{3} +\thisrow{4} +\thisrow{5}
 }
]{sum}{\dataA}

\pgfplotsset{
  percentage plot/.style={
    point meta=explicit,
    every node near coord/.append style={
      font=\tiny,
      color=black,
    },
    nodes near coords={
      \pgfmathtruncatemacro\iszero{\originalvalue==0}
      \ifnum\iszero=0
      \pgfmathprintnumber[fixed,fixed zerofill,precision=1]{\pgfplotspointmeta}
      \fi
    },
    yticklabel=\pgfmathprintnumber{\tick}\,$\%$,
    ymin=0,
    ymax=100.01, 
    visualization depends on={y \as \originalvalue},
    enlarge x limits={abs=10mm}
  },
  percentage series/.style={
    table/x expr=\coordindex, 
    table/y expr=(\thisrow{#1}/\thisrow{sum}*100),
    table/meta=#1
    }
}
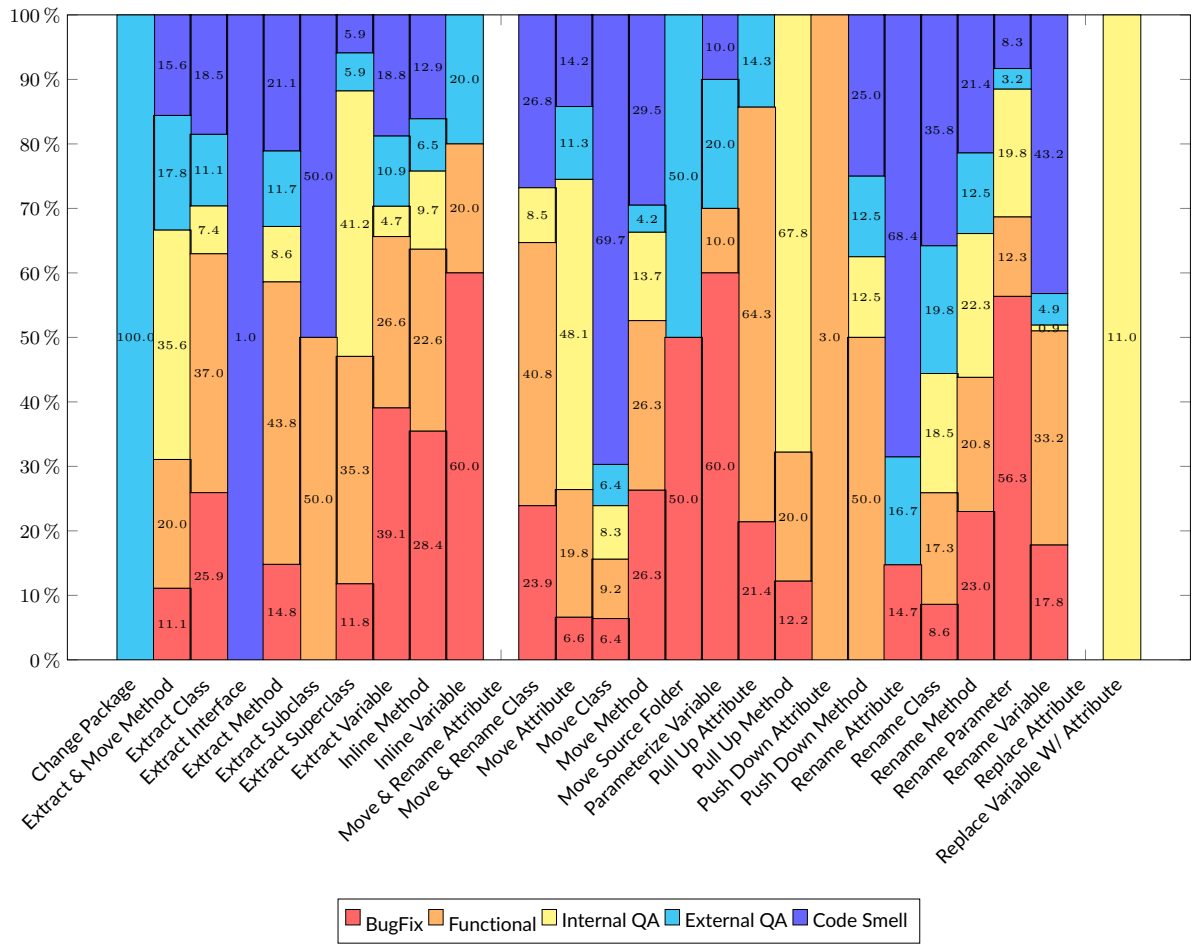
\begin{figure*}[h]  
\centering 
\begin{tikzpicture}
\begin{scope}[scale=0.9]
\begin{axis}[
    ybar stacked,
    height=11cm,
    width=18cm,
    percentage plot,
    bar width=0.55cm, 
    xticklabels from table={\dataA}{Motivation}, 
    xtick=data,
    x tick label style={
      rotate=45,
      anchor=east, 
      xshift=-1.5mm, yshift=-2mm
    },
    legend style={
      at={(0.5,-0.44)},
      anchor=south,
      legend columns=-1
      },
]

    \addplot [fill=red!60]  table[percentage series=1] {\dataA};
    \addplot [fill=orange!60]                 table[percentage series=2] {\dataA};
    \addplot [fill=yellow!60]       table[percentage series=3] {\dataA};
    \addplot [fill=cyan!60]       table[percentage series=4] {\dataA};
    \addplot [fill=blue!60]       table[percentage series=5] {\dataA};


    \legend{\strut BugFix, \strut Functional, Internal QA, External QA, Code Smell}
\end{axis}
\end{scope}
\end{tikzpicture}
\caption{Distribution of refactoring types per category performed by the rest of developers}
\label{fig:ref_dist_class_rest}
\end{figure*}


\begin{figure*}[htbp]
	\centering
    \includegraphics[scale = 0.35]{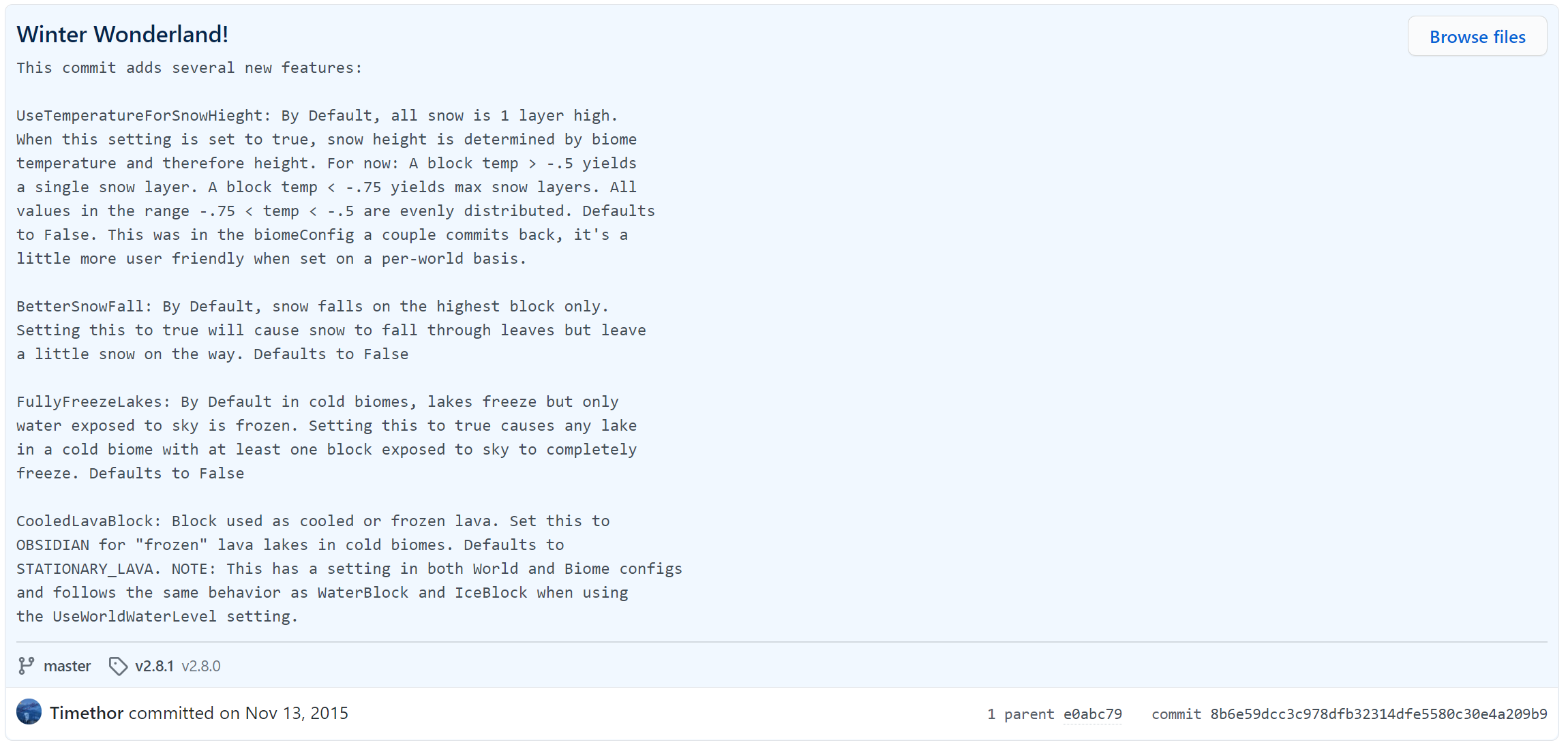}   
    \caption{Commit message indicating the addition to a new feature \cite{functional}}
    \label{fig:CommitMessageExample-functional}

\vspace{0.70cm}

	\centering
    \includegraphics[scale = 0.35]{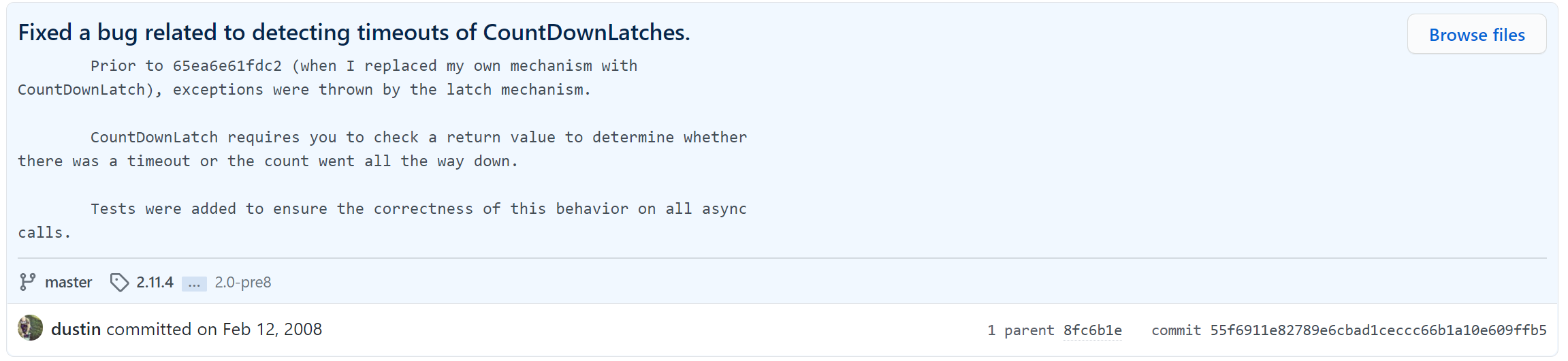}   
    \caption{Commit message indicating the fix of the bug \cite{bugfix}}
    \label{fig:CommitMessageExample-bugfix}

\vspace{0.70cm}

\centering 
\includegraphics[scale = 0.35]{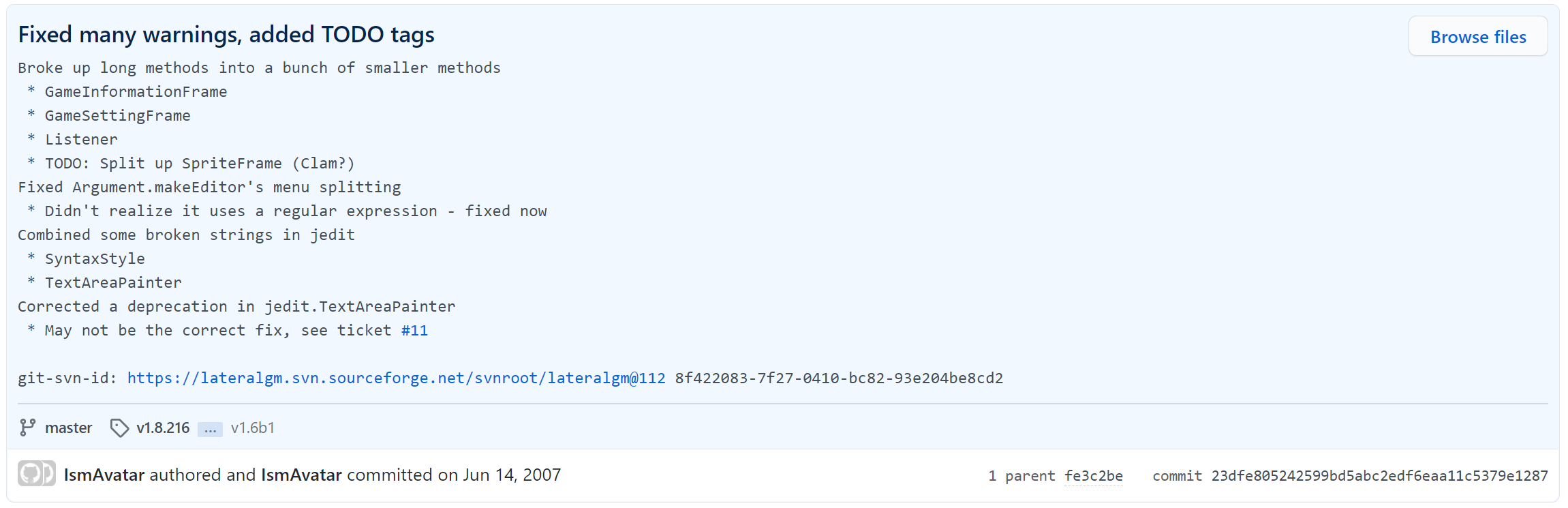}
\caption{Commit message indicating the removal of the code smell \cite{codesmell}}
\label{fig:CommitMessageExample-codesmell}

\vspace{0.70cm}

\centering 
\includegraphics[scale = 0.35]{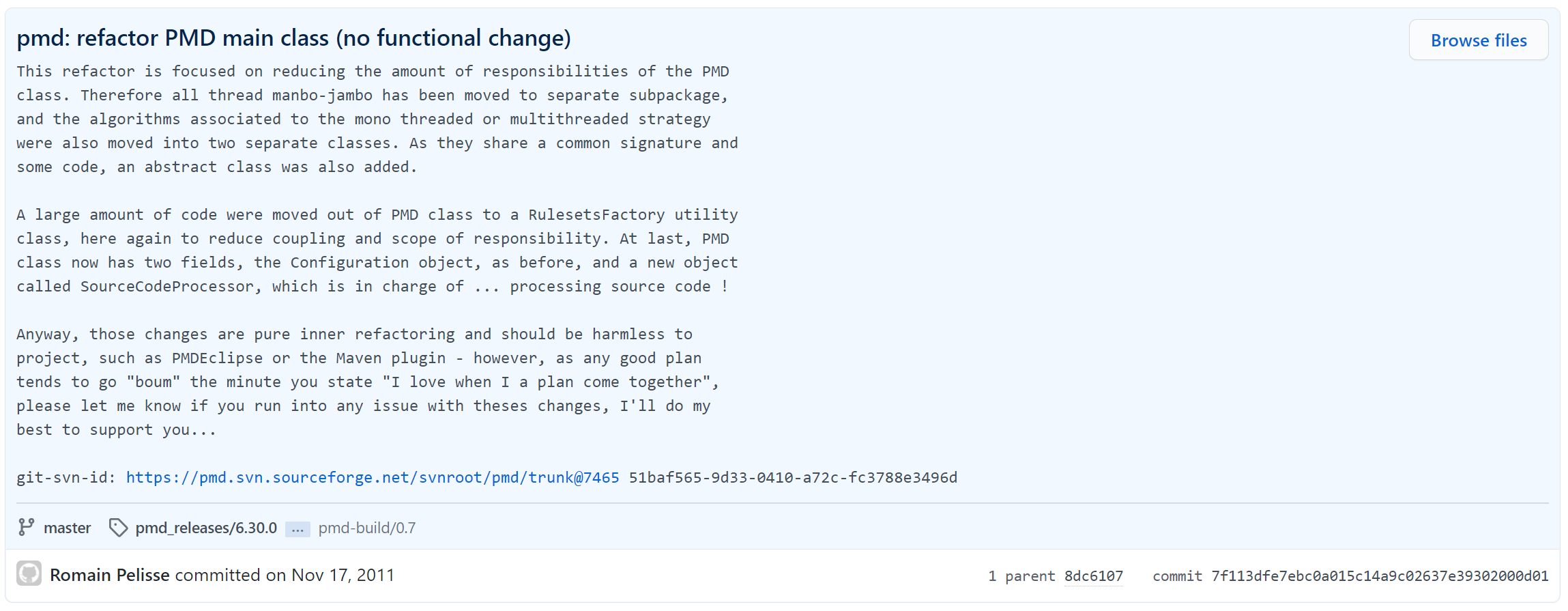}
\caption{Commit message indicating the improvement of the internal quality attribute \cite{internal}}
\label{fig:CommitMessageExample-internal}

\vspace{0.70cm}

\centering 
\includegraphics[scale = 0.35]{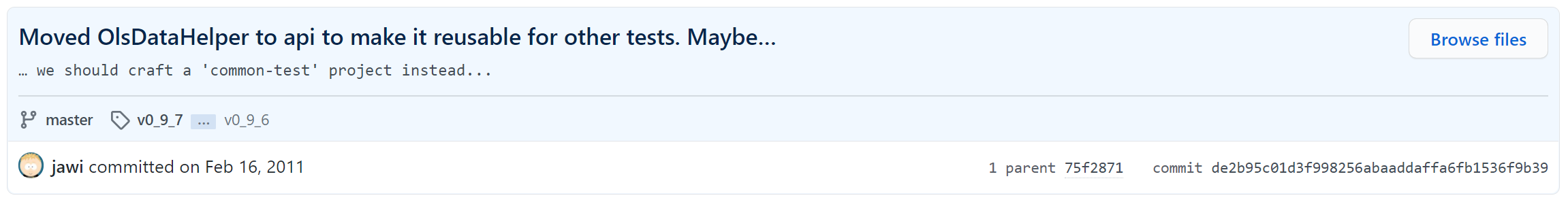}
\caption{Commit message indicating the improvement of the external quality attribute \cite{external}}
\label{fig:CommitMessageExample-external}
\end{figure*}

To better understand the nature of classified commits, we randomly sampled examples from each category 
to illustrate the type of information contained in these messages, and how it infers the use of refactorings in specific contexts.
Table \ref{Table:ClassificationCategories} shows the five main motivations driving refactoring, which we can also divide into more fine-grained subcategories. For instance, we subcategorize \textit{Functional}-classified commits into \textit{addition}, \textit{update} and \textit{deletion}. Similarly, \textit{BugFix} is decomposed into \textit{Localization}, \textit{debugging} and \textit{correction}. The corresponding subcategories for \textit{Code Smell Resolution} include \textit{long method}, \textit{duplicate code removal} and \textit{large class}. As for the internal, the subcategories could include Object-Oriented design improvement such as \textit{coupling} and \textit{cohesion}. 
The subcategories of \textit{External Quality Attribute} are more straightforward to extract from the commit messages since developers tend to explicitly mention which quality attribute they are trying to optimize. For this study, the sub-categories we found in our mined commits include \textit{testability}, \textit{usability}, \textit{performance}, \textit{reusability} and \textit{readability}. Then, for each subcategory, we provide an illustrative commit message as an example as shown in Fgures \ref{fig:CommitMessageExample-functional}, \ref{fig:CommitMessageExample-bugfix}, \ref{fig:CommitMessageExample-codesmell}, \ref{fig:CommitMessageExample-internal}, and \ref{fig:CommitMessageExample-external}. One important observation that can be drawn from these messages is that developers may have multiple reasons to refactor the code; some of which go beyond Martin Fowler's traditional definition of associating refactoring with improving design by removing code smells. These subcategories are not exhaustive, there are many others. These are just examples to illustrate what commits look like.

\begin{itemize}
 \item For functional-focused refactoring, developers do refactor the code to introduce, modify, or delete features. For instance, the commit description in one of the analyzed commits (see Figure \ref{fig:CommitMessageExample-functional}) was: \textit{adds several new features: UseTemperatureForSnowHieght, etc}. It is clear that the intention of the refactoring was to enhance existing functionalities related to the snow measurement, which include using temperature for snow height, changing the freeze setting, etc. As can be seen, developers incorporate refactoring activities in development-related task (i.e., feature addition).
 \item In the second category, developers perform refactoring to facilitate bug fix-related activities: resolution, debugging, and localization. As described in the following commit comment (See Figure \ref{fig:CommitMessageExample-bugfix}): \textit{Fixed a bug related to detecting timeouts of CountDownLatches, etc}, the developer who performed refactoring explained that the purpose for the refactoring was to resolve certain bug that required to check a return value to determine whether there was a timeout or the count went all the way down. Thus, it is clear that developers frequently floss refactor since they intersperse refactoring with other programming activity (i.e., bug fixing).
 \item For code smell-focused refactoring, it is clear that developers remove certain code smells such as feature envy, duplicated code, and long methods. As can be seen from from Figure \ref{fig:CommitMessageExample-codesmell}, developers performed \textit{Extract Method} refactoring to remove a code smell which corresponds to a long method bad smell. This traditional design improvement refactoring motivation is best illustrated in the following change message: \textit{Broke up long methods into a bunch of smaller methods}.
 \item To improve the internal design, it is apparent that developers introduce good practices (e.g., use inheritance, polymorphism, and enhance the main modularization quality drivers). Further, developers primarily refactor the code to improve the dominant modularization driving forces (i.e., cohesion and coupling) to maximize intra-class connectivity and minimize inter-class connectivity. This design improvement refactoring motivation is best illustrated in the following change message, shown in Figure~\ref{fig:CommitMessageExample-internal}: \textit{A large amount of code were moved out of PMD class to a RulesetsFactory utility
class, here again to reduce coupling and scope of responsibility.}
 \item For the external quality attribute category, refactorings are performed to enhance nonfunctional attributes. For example, developers refactor the code to improve its testability, usability, performance, reusability, and readability. Developers are making changes such as extracting a method for improving testability as they test parts of the code separately. They are also moving a method or class to improve code reusability. This is illustrated by the following commit message in Figure \ref{fig:CommitMessageExample-external}: \textit{Moved OlsDataHelper to api to make it reusable for other tests.} Closer inspection of this commit comment show that developers intended to apply nonfunctional-related development topics while performing refactorings.
 
\end{itemize}

From the refactoring operation usage perspective, we notice that some commit messages describe the method of refactoring identified by Refactoring Miner. For instance, in order to remove the long method code smell, developers extracted a method to reduce the length of the original method body. Also, to remove the feature envy code smell, move method refactoring operation was performed in order to place the fields and methods in their preferred class. In both cases, developers explained these changes in the commit messages. It is worth noting that a singular refactoring is almost never performed on its own, as was noted for some of the refactoring commit messages reviewed for this paper. For instance, to eliminate a long method, Fowler suggests using several refactorings (e.g., \textit{Replace Temp with Query}, \textit{Introduce Parameter Object}, and \textit{Preserve Whole Object})  depending on the complexity of the transformation. Due to the limited refactoring operations supported by Refactoring Miner, we could not demonstrate such composite refactorings. However, exploring developers practice in performing batch/composite refactoring is an interesting research direction that can be investigated in the future.

To better analyze the existence of any patterns on the types of refactorings applied in each category, Figures~\ref{fig:ref_dist_class_top5developers} and \ref{fig:ref_dist_class_rest} present the distribution of refactoring types in every category. We would like to note that these two stacked bar charts normalized the values to 100\%, i.e., the charts show the contribution for each group toward the 100\%. Refactoring types were applied by top 5\% developers with similar frequency across all categories. However, the most used refactoring types for Bug Fix, Functional, Internal Quality Attribute, External Quality Attribute, and Code Smell are respectively: \textit{Extract Variable}, \textit{Move \& Rename Attribute}, \textit{Replace Attribute}, \textit{Move Source Folder}, and \textit{Move Class}. In contrast, as can be seen from Figure \ref{fig:ref_dist_class_rest}, the remaining set of developers perform refactorings differently as not all refactoring types were applied in all categories and other types were not applied in any of the categories such as \textit{Move \& Rename Attribute} and \textit{Replace Attribute}. What can be clearly seen in this figure is the variability of certain refactoring types and frequency. For example, \textit{Change Package}, \textit{Extract Interface}, \textit{Push Down Attribute} and \textit{Replace Variable with Attribute} are solely applied in External Quality Attribute, Code Smell, Functional, and Internal Quality Attribute, respectively. Similarly, \textit{Move Class} was mostly used by the rest of developers when refactoring code to fix code smells.

We compare the distribution of refactoring refactorings identified for each category by the top 5\% developers and the remaining set of refactoring contributors using the Wilcoxon sum-rank test \cite{conover1998practical}; a pairwise statistical test verifying whether two sets have a similar distribution. The Null hypothesis is defined by no variation in the refactoring distribution performed by top 5\% developers and the rest of developers. The alternative hypothesis indicates that there is a variation in the refactoring distribution per category between both group of developers. If the p-value is smaller than 0.05, the distribution difference between the two sets is considered statistically significant. The choice of Wilcoxon comes from its non-parametric nature with no assumption of a normal data distribution. The difference between the contribution of the two groups of developers are found to be statistically significant as the \textit{p}-values for the category BugFix, Functional, Internal QA, External QA, and Code Smell are respectively 3.17724e-8, 4.05194e-9, 3.76671e-10, 6.69847e-9, and 8.81574e-8.

\vspace{.2cm}
\begin{tcolorbox}
\textit{Summary}. Our classification has shown that improving the design through fixing code smells is not the main driver for top contributed and less contributed developers to refactoring their code bases. As explicitly mentioned by the developers in their commit messages, refactorings are interleaved with the activities of bugfix and feature addition or modification. Additionally, the distribution of refactoring operations per refactoring motivation are performed differently by the top 5\% and the remaining set of developers.
\end{tcolorbox}

\subsection{\RQD}
\begin{table*}[htbp]
\centering
\caption{List of Self-Affirmed Refacoring (SAR) patterns defined in \cite{alomar2020howwe}}
\label{Table:GeneralPatterns}
\begin{sideways}
\small
\resizebox{\textwidth}{!}{%
\begin{tabular}{lllll}
\toprule
\textbf{Patterns}\\
\midrule
(1) Add* & (47) Chang* & (93) Cleaned out & (139) CleanUp & (185) CleaningUp \\ 
(2) Clean* up & (48) Clean-up & (94) Creat* & (140) Decompos* & (186)  Encapsulat*  \\ 
(3) Enhanc*  & (49) Extend* & (95) Extract* & (141) Factor* Out & (187) Fix*   \\ 
(4) Improv* & (50) Inlin* & (96) Introduc* & (142) Merg* & (188) Migrat*  \\ 
(5) Modif*  & (51) Modulariz* &  (97) Mov* & (143) Organiz* & (189) Polish*  \\ 
(6) Pull* Up  & (52) PullUp  & (98) Push Down & (144) PushDown & (190) Repackag*   \\ 
 (7) Re packag* & (53) Re-packag* & (99) Redesign*  & (145) Re-design* & (191) Reduc* \\ 
(8) Refactor* & (54) Refin* & (100) Reformat* & (146) Remov*  & (192) Renam*  \\ 
(9) Reorder* & (55) Reorganiz* & (101) Re-organiz*  & (147) Repackag* & (193) Replac*   \\ 
(10) Restructur*& (56) Rework* & (102) Rewrit*  & (148) Re-writ* & (194) Rewrot*    \\ 
(11) Simplif*& (57) Split* & (103) TidyUp & (149) Tid*-up & (195) Tid* Up   \\ 
         (12) A bit of refactor* & (58) Basic code clean up & (104) Chang* code style & (150) Ease maintenance moving forward & (196) Replace it with \\ 
         (13) Big refactor*  & (59) Big cleanup & (105) Clean* up the code style & (151) Ease of code maintenance & (197) Extracted out code \\ 
         (14) Better factored code & (60) Cleanliness & (106) Code style improv* & (152) Easier to maintain & (198) Reduced code dependency\\ 
         (15) Code refactor* & (61) Clean* up unnecessary code & (107) Code style unification & (153) Simplify future maintenance & (199) Pushed down dependencies  \\ 
         (16) Code has been refactored extensively & (62) Cleanup formatting  &  (108) Fix code style & (154) Improve quality &  (200) Simplify the code\\
         (17) Extensive refactor* & (63) Code clean & (109) Improv* code style &  (155) Improvement of code quality & (201) Less code    \\ 
         (18) Refactoring towards nicer name analysis & (64) Code cleanup  & (110) Minor adjustments to code style & (156) Improved style and code quality &  (202) Change package\\ 
         (19) Heavily refactored code & (65) Code cleanliness & (111) Modifications to code style & (157) Maintain quality & (203) Cosmetic changes \\ 
         (20) Heavy refactor* & (66) Code clean up & (112) Lots of modifications to code style &  (158) More quality cleanup & (204) Full customization  \\      
         (21) Little refactor* & (67) Massive cleanup  & (113) Makes the code easier to program   & (159) Better name & (205) Structure change \\ 
         (22) Lot of refactor* & (68) Minor cleaning of the code   & (114) Code review & (160) Chang* name & (206) Module structure change  \\
         (23) Major refactor* &  (69) Housekeeping  & (115) Code rewrite & (161) Chang* the name & (207) Module organization structure change\\ 
         (24) Massive refactor*  & (70) Major rewrite and simplification & (116) Code cosmetic & (162) Chang* the package name & (208) Polishing code\\ 
         (25) Huge refactor* & (71) Improv* consistency & (117) Code revision & (163) Chang* method name & (209) Improv* code quality \\ 
         (26) Minor refactor* & (72) Some fix* and optimization  & (118) Code optimization & (164) Chang* method parameter names for consistency & (210) Chang* package structure\\ 
         (27) More refactor* & (73) Minors fix* and tweak & (119) Code reformatting & (165) Enables condensed naming & (211) Fix quality flaws
\\ 
         (28) Refactor* code & (74) Fix* annoying typo  &  (120) Code organization &  (166) Fix* naming convention & (212) Get rid of\\ 
         (29) Refactor* existing schema & (75) Fix* some formatting & (156) Code rearrangement &  (167) Fix nam* & (213) hang* design\\ 
         (30) Refactor out & (76) Fix* formatting    & (122) Code formatting & (168) Typo in method name & (214) Improv* naming consistency  \\ 
         (31) Small refactor* & (77) Modifications to make it work better  & (123) Code polishing & (169) Maintain convention & (215) Remov* unused classes \\ 
         (32) Some refactor* & (78) Make it simpler to extend  & (124) Code simplification & (170) Maintain naming consistency & (216) Minor simplification  \\ 
         (33) Tactical refactor* & (79) Fix* Regression & (125) Code adjustment &  (171) Major renam* & (217) Fix* quality issue \\ 
         (34) Moved a lot of stuff  & (80) Remov* the useless  & (126)  Code improvement & (172) Name cleanup & (218) Naming improvement  \\ 
         (35) Fix this tidily  & (81) Remov* unneeded variables & (127) Code style & (173) Renam* for consistency & (219) Packaging improvement\\ 
         (36) Further tidying  & (82) Remov* unneeded code   & (128) Code restructur* & (174) Renam* according to java naming conventions  & (220) Structural chang* \\ 
         (37) Tidied up and tweaked &  (83) Remov* redundant & (129) Code beautifying  & (175) Renam* classes for consistency & (221) Hierarchy clean* \\ 
         (38) Tidied up some code & (84) Remov* dependency  & (130) Code tidying & (176) Renam* package & (222) Hierarchy reduction\\
          (39) Restructur* package & (85) Remov* unused dependencies & (131) Code enhancement  & (177) Resolv* naming inconsistency & (223) Enhanc* architecture \\ 
        (40) Restructur* code &  (86) Remov* unused  & (132) Code reshuffling &  (178) Simpler name & (224) Architecture enhanc*\\ 
         (41) Aggregat* code & (87) Remov* unnecessary else blocks & (133) Code modification  & (179) Us* appropriate variable names & (225) Trim unneeded code\\ 
         (42) Beautif* code  & (88) Remov* needless loop   & (134) Code unification & (180) Us* more consistent variable names & (226) Remov* unneeded code   \\ 
         (43) Tidy code & (89) Maintain consistency & (135) Code quality & (181) Neaten up & (227) More consistent formatting  \\ 
         (44) Beautify* & (90) Customiz* & (136) Make code clearer & (182) Moved more code out of & (228) More easily extended \\
         (45) Moved all integration code to separate package & (91) Improve code clarity  & (137) Code clarity  & (183) Fix bad merge & (229) Makes it more extensible-friendly \\ 
         (46) Improve code & (92) Simplify code & (138) Clean* code & (184) Cleanup code & (230) Clean* up code \\ \hline
       
\end{tabular}}
\end{sideways}
\end{table*} 

Generally, having good historical documentation is invaluable when tracking the root cause of a regression or bug, and having a good refactoring documentation help to better understand the motivation driving refactoring. Giving enough documentation/background related to the performed refactorings is important to facilitate the code review process. Recent studies have shown that the process of reviewing refactoring changes heavily relies on understanding the context of the performed refactoring \cite{alomar2021xerox,paixao2020behind}. Furthermore, the lack of refactoring documentation was one of the main challenges that impacted the efficiency of the review process \cite{alomar2021xerox}. This observation motivates us to explore documentation practice written by open-source developers and investigate the correlation between developer contributions and refactoring documentation.

In order to better understand developers contribution and its relation to the practice of refactoring documentation, we propose the following hypothesis: \say{\textit{top contributed developers tend to document refactoring activities less than less contributed developers}}. To test this hypothesis, we extract all of the commit messages mined by the tool Refactoring Miner, and consider Self-Affirmed Refactoring (SAR) terminology patterns listed in \cite{alomar2019can,alomar2020toward,alomar2020howwe} to observe refactoring documentation practice expressed by the top 5\% and the remaining set of developers in their commit messages.

\pgfplotsset{compat=1.14}
\definecolor{findOptimalPartition}{HTML}{696969}
\definecolor{storeClusterComponent}{HTML}{808080}
\definecolor{dbscan}{HTML}{BEBEBE}
\definecolor{constructCluster}{HTML}{DCDCDC} 

\pgfplotstableread[col sep=comma,header=true]{
Top,1,2
Change Package, 100, 0
Extract \& Move Method, 35.0, 65.0
Extract Class, 29.5, 70.5
Extract Interface, 100, 0
Extract Method, 26.6, 73.4
Extract Subclass, 25.0, 75.0
Extract Superclass, 20.3, 79.7
Extract Variable, 19.6, 80.4
Inline Method, 3, 67.3
Inline Variable, 100, 0
Move \& Rename Attribute, 100, 0
Move \& Rename Class, 57.2, 42.8
Move Attribute, 33.2, 66.8
Move Class, 64.2, 35.8
Move Method, 40.1, 59.9
Move Source Folder, 10.7, 89.3
Parameterize Variable, 16.7, 83.3
Pull Up Attribute, 43.2, 56.8
Pull Up Method, 30.0, 70.0
Push Down Attribute, 100, 0
Push Down Method, 21.1, 78.9
Rename Attribute, 24.7, 75.3
Rename Class, 37.9, 62.1
Rename Method, 38.0, 62.0
Rename Parameter, 32.6, 67.4
Rename Variable, 33.5, 66.5
Replace Attribute, 100, 0
Replace Variable W/ Attribute, 100, 0

}\dataB

\pgfplotstablecreatecol[
 create col/expr={
   \thisrow{1} + \thisrow{2}
 }
]{sum}{\dataB}

\pgfplotsset{
  percentage plot/.style={
    point meta=explicit,
    every node near coord/.append style={
      font=\tiny,
      color=black,
    },
    nodes near coords={
      \pgfmathtruncatemacro\iszero{\originalvalue==0}
      \ifnum\iszero=0
      \pgfmathprintnumber[fixed,fixed zerofill,precision=1]{\pgfplotspointmeta}
      \fi
    },
    yticklabel=\pgfmathprintnumber{\tick}\,$\%$,
    ymin=0,
    ymax=100.01, 
    visualization depends on={y \as \originalvalue},
    enlarge x limits={abs=10mm}
  },
  percentage series/.style={
    table/x expr=\coordindex, 
    table/y expr=(\thisrow{#1}/\thisrow{sum}*100),
    table/meta=#1
    }
}


\begin{figure}[h]  
\centering 
\begin{tikzpicture}
\begin{scope}[scale=0.9]
\begin{axis}[
    ybar stacked,
    height=11cm,
    width=18cm,
    percentage plot,
    bar width=0.55cm, 
    xticklabels from table={\dataB}{Top}, 
    xtick=data,
    x tick label style={
      rotate=45,
      anchor=east, 
      xshift=-1.5mm, yshift=-2mm
    },
    legend style={
      at={(0.5,-0.44)},
      anchor=south,
      legend columns=-1
      },
]

    \addplot [fill=pink] table[percentage series=1] {\dataB};
    \addplot [fill=constructCluster] table[percentage series=2] {\dataB};


    \legend{\strut Top 5\% developers, \strut Rest of developers}
\end{axis}
\end{scope}
\end{tikzpicture}
\caption{Distribution of refactoring commits labeled with the keyword `Refactor' performed by the top 5 \% and the remaining set of developers}
\label{fig:ref_refactor_top5rest}
\end{figure}
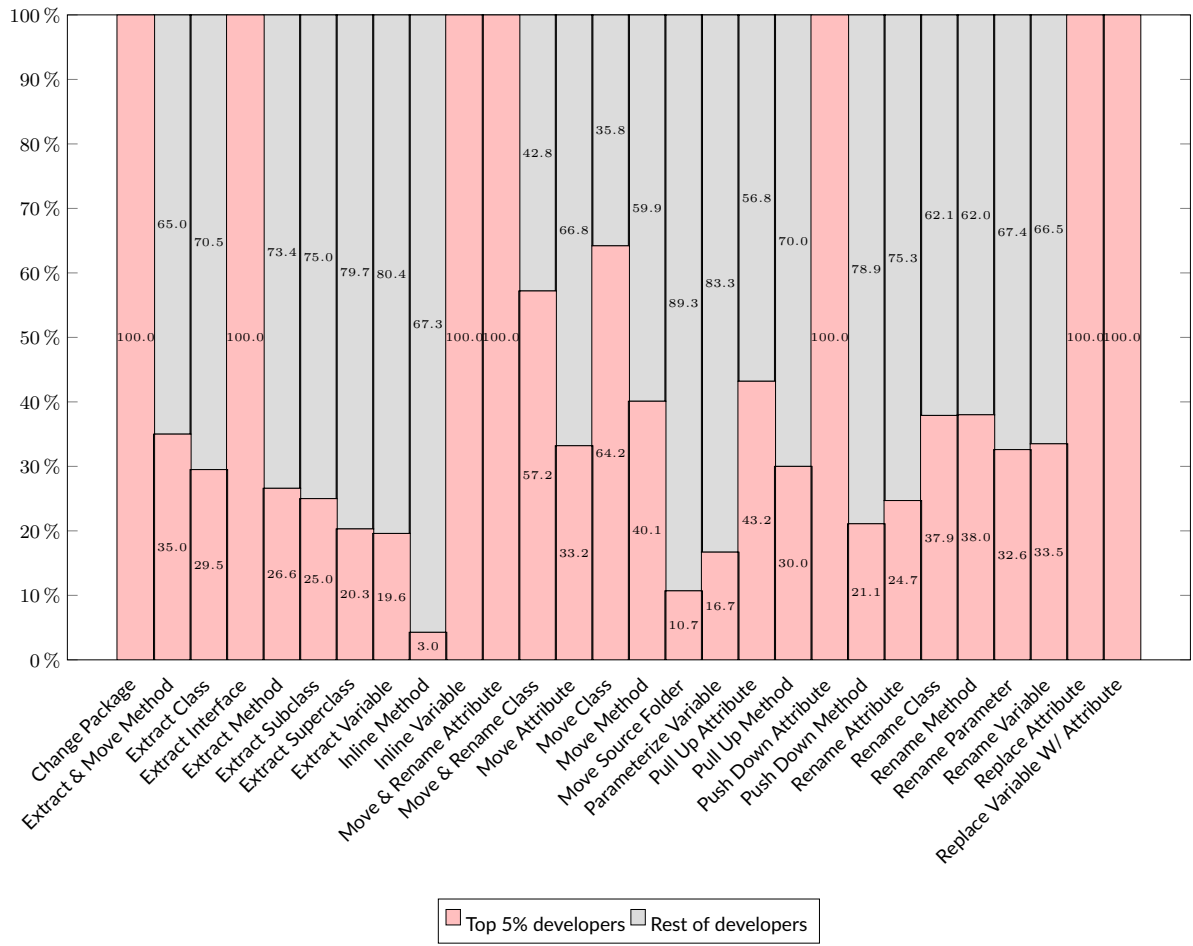

\pgfplotsset{compat=1.14}
\definecolor{findOptimalPartition}{HTML}{696969}
\definecolor{storeClusterComponent}{HTML}{808080}
\definecolor{dbscan}{HTML}{BEBEBE}
\definecolor{constructCluster}{HTML}{DCDCDC} 

\pgfplotstableread[col sep=comma,header=true]{
Rest,1,2
Change Package, 48.0, 52.0
Extract \& Move Method, 48.6, 51.4
Extract Class, 46.9, 53.1
Extract Interface, 100, 0
Extract Method, 48.0, 52.0
Extract Subclass, 47.6, 52.4
Extract Superclass, 50.8, 49.2
Extract Variable, 47.1, 52.9
Inline Method, 46.6, 53.4
Inline Variable, 46.5, 53.4
Move \& Rename Attribute, 100, 0
Move \& Rename Class, 51.2, 48.8
Move Attribute, 46.7, 53.3
Move Class, 50.3, 49.7
Move Method, 47.7, 52.3
Move Source Folder, 47.8, 52.2
Parameterize Variable, 45.7, 54.3
Pull Up Attribute, 46.8, 53.2
Pull Up Method, 50.4, 49.6
Push Down Attribute, 47.2, 52.8
Push Down Method, 54.4, 45.6
Rename Attribute, 49.1, 50.9
Rename Class, 49.1, 50.9
Rename Method, 49.7, 50.3
Rename Parameter, 48.9, 51.1
Rename Variable, 50.2, 48.9
Replace Attribute, 46.0, 54.0
Replace Variable W/ Attribute, 100, 0

}\dataB

\pgfplotstablecreatecol[
 create col/expr={
   \thisrow{1} + \thisrow{2}
 }
]{sum}{\dataB}

\pgfplotsset{
  percentage plot/.style={
    point meta=explicit,
    every node near coord/.append style={
      font=\tiny,
      color=black,
    },
    nodes near coords={
      \pgfmathtruncatemacro\iszero{\originalvalue==0}
      \ifnum\iszero=0
      \pgfmathprintnumber[fixed,fixed zerofill,precision=1]{\pgfplotspointmeta}
      \fi
    },
    yticklabel=\pgfmathprintnumber{\tick}\,$\%$,
    ymin=0,
    ymax=100.01, 
    visualization depends on={y \as \originalvalue},
    enlarge x limits={abs=10mm}
  },
  percentage series/.style={
    table/x expr=\coordindex, 
    table/y expr=(\thisrow{#1}/\thisrow{sum}*100),
    table/meta=#1
    }
}

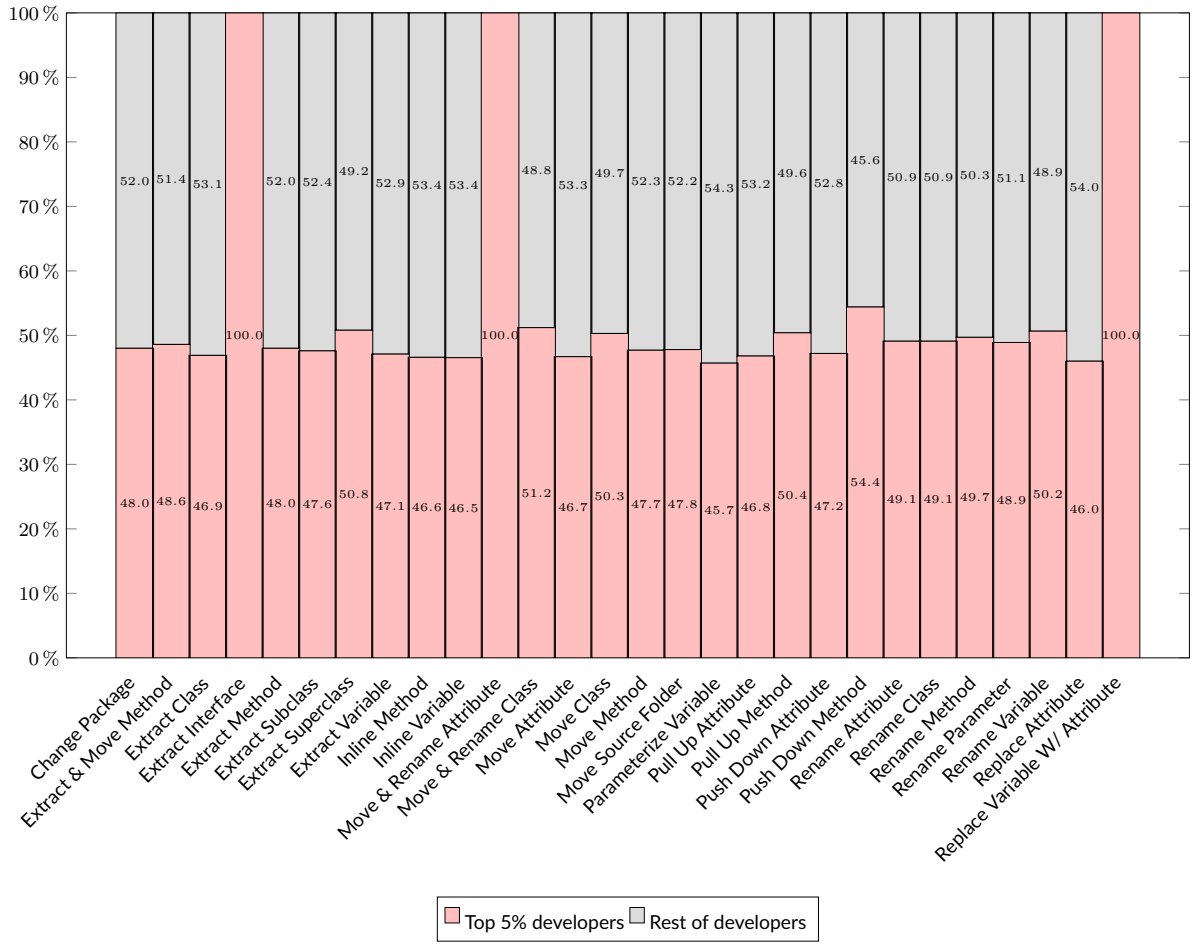
\begin{figure}[h]  
\centering 
\begin{tikzpicture}
\begin{scope}[scale=0.9]
\begin{axis}[
    ybar stacked,
    height=11cm,
    width=18cm,
    percentage plot,
    bar width=0.55cm, 
    xticklabels from table={\dataB}{Rest}, 
    xtick=data,
    x tick label style={
      rotate=45,
      anchor=east, 
      xshift=-1.5mm, yshift=-2mm
    },
    legend style={
      at={(0.5,-0.44)},
      anchor=south,
      legend columns=-1
      },
]

    \addplot [fill=pink] table[percentage series=1] {\dataB};
    \addplot [fill=constructCluster] table[percentage series=2] {\dataB};


    \legend{\strut Top 5\% developers, \strut Rest of developers}
\end{axis}
\end{scope}
\end{tikzpicture}
\caption{Distribution of refactoring commits labeled with the SAR patterns performed by the top 5 \% and the remaining set of developers}
\label{fig:ref_sar_top5rest}
\end{figure}
\pgfplotsset{compat=1.14}
\definecolor{findOptimalPartition}{HTML}{696969}
\definecolor{storeClusterComponent}{HTML}{808080}
\definecolor{dbscan}{HTML}{BEBEBE}
\definecolor{constructCluster}{HTML}{DCDCDC} 

\pgfplotstableread[col sep=comma,header=true]{
Top,1,2
Change Package, 100, 0
Extract \& Move Method, 98.6, 1.4
Extract Class, 97.4, 2.6
Extract Interface, 100, 0
Extract Method, 98.5, 1.5
Extract Subclass, 98.9, 1.1
Extract Superclass, 98.1, 1.9
Extract Variable, 96.2, 3.8
Inline Method, 93.9, 6.1
Inline Variable, 100, 0
Move \& Rename Attribute, 100, 0
Move \& Rename Class, 99.3, 0.7
Move Attribute, 99.0, 1.0
Move Class, 99.9, 0
Move Method, 97.9, 2.1
Move Source Folder, 98.9, 1.1
Parameterize Variable, 97.4, 2.6
Pull Up Attribute, 99.7, 0
Pull Up Method, 99.1, 0.9
Push Down Attribute, 100, 0
Push Down Method, 99.6, 0.4
Rename Attribute, 97.8, 2.2
Rename Class, 98.5, 1.5
Rename Method, 92.0, 8.0
Rename Parameter, 93.0, 7.0
Rename Variable, 98.6, 1.4
Replace Attribute, 100, 0
Replace Variable W/ Attribute, 100, 0

}\dataB

\pgfplotstablecreatecol[
 create col/expr={
   \thisrow{1} + \thisrow{2}
 }
]{sum}{\dataB}

\pgfplotsset{
  percentage plot/.style={
    point meta=explicit,
    every node near coord/.append style={
      font=\tiny,
      color=black,
    },
    nodes near coords={
      \pgfmathtruncatemacro\iszero{\originalvalue==0}
      \ifnum\iszero=0
      \pgfmathprintnumber[fixed,fixed zerofill,precision=1]{\pgfplotspointmeta}
      \fi
    },
    yticklabel=\pgfmathprintnumber{\tick}\,$\%$,
    ymin=0,
    ymax=100.01, 
    visualization depends on={y \as \originalvalue},
    enlarge x limits={abs=10mm}
  },
  percentage series/.style={
    table/x expr=\coordindex, 
    table/y expr=(\thisrow{#1}/\thisrow{sum}*100),
    table/meta=#1
    }
}

In an empirical context, we test this hypothesis in two rounds. In the first round, we used the term `refactor' since it is used by the related studies \citep{kim2014empirical,Ratzinger:2008:RRS:1370750.1370759,zhangpreliminary18,murphy2012we,soares2013comparing} and intuitively the first term to identify refactoring-related activities in commit messages. In the second round, we re-tested the hypothesis using the SAR patterns (See Table \ref{Table:GeneralPatterns}). For both rounds, we quantified the proportion of commit messages including the searched label for top 5 \% and the rest of refactoring contributors.

Figures \ref{fig:ref_refactor_top5rest} and \ref{fig:ref_sar_top5rest} portray the distribution of refactorings in labeled commits with `refactor' and SAR patterns, respectively. Similar to the previous research question, we used 100\% stacked bar charts to easily compare and visualize the difference between both groups. The first observation we can draw is that top contributed developers tend to document less refactoring than less contributed developers. Since these developers frequently refactor the code, they have less time to document. Another reason can be that top contributed developers feel less need to document refactoring activities because the changes are clear and easy to understand. More research is required to explore and better understand this phenomena.  
In contrast, developers who occasionally refactor the code, they tend to provide better refactoring documentation through commit messages. 
Another observation is that top contributed developers tend to follow the name of the refactoring operations mentioned in the Fowler’s book \cite{Fowler:1999:RID:311424}. For example, to perform rename-related activities, they use the term `Rename' in the corresponding commit messages.

To determine whether the variation is statistically significant, we use the Wilcoxon sum-rank test \cite{conover1998practical}, a non-parametric test, to compare between the two group of commits, since these groups are independent on one another. The Null hypothesis is defined by no variation in the refactoring distribution performed by top 5\% developers and the remaining of the contributors. Thus, the alternative hypothesis indicates that there is a variation in the usage of patterns between both sets. The variation between values of both sets is considered significant if its associated p-value is less than 0.05. By comparing the different commits that are labeled `refactor' and labeled with SAR patterns by top 5 \% developers and the rest of developers, we observe a significant number of labeled refactoring commits using the term `refactor' for each refactoring operation supported by the tool Refactoring Miner (\textit{p}-value = 0.04). The results for commits labeled with SAR patterns is statistically significant (\textit{p}-value = 0.00009)
This implies that there is a strong trend of less contributed developers in using these phrases in refactoring commits.


\begin{tcolorbox}
\textit{Summary}. 
Refactoring contributors that frequently refactor the code tend to document refactoring less than developers that occasionally perform refactoring. Our conjecture is that top refactoring contributors found that the changes are clear, and so, generic expression of the refactoring would be sufficient, in contrast to the rest of developers that provide descriptive refactoring documentation to the performed refactoring activity.
\end{tcolorbox}

\section{Research implications and future directions} \label{section:implication}


The main implications of this study are as follows:
\begin{itemize}
    \item \textbf{Improving the design through fixing code smells is not the main driver for top contributed and less contributed developers to refactor their code bases.} Our RQ3 finding has major implications for facilitating refactoring support for developers. It may be necessary for future tools to motivate developers not only by pointing out how they can refactor code smells, but how the refactoring will help them from a multi-faceted point of view, i.e., what are all of the characteristics that this refactoring will improve? Further, this implication makes it clear that we need to study the context around different refactorings to understand why developers perform them so that recommendation systems can be made to mimic this reasoning whenever it is found that the reasoning is based on a solid foundation.
    
    \item{ \textbf{Encouraging continuous refactoring as part of making code changes.} Refactoring is an integral part of the software development process. To increase the efficiency of refactoring techniques, it is important to know when and who should refactor the code. Our finding  shows that top refactoring contributors perform the majority of refactoring. To improve the state of practice of refactoring, the refactoring tasks should be distributed evenly between developers. Further, it is noticeable from our RQ3 finding that developers interleave refactoring with other development-related tasks (e.g., add features and fix bugs). Since the impact of this activity on quality of the code is unclear, future developers are encouraged not to create new features during the refactoring process. Instead, it is better to refactor the code before updating the code. In their study of refactoring practice in modern code review at Xerox, AlOmar et al. \cite{alomar2021xerox} found that participants acknowledged that mixing refactoring with any other activity is a potential problem as the behavior preservation cannot be guaranteed and this task might introduce new bugs. Moreover,  a recent study \cite{alomar2021SLM} shows there are different approaches to test if the transformation is behaviorally preserved. We can now encourage developers with all levels of expertise to test the performed refactoring using the suitable approach.}
    
    \item \textbf{Improving the quality of refactoring documentation.} Our study helps us understand refactoring documentation practices that trigger the need to explore the motivation behind refactoring. The study helps future developers to follow best documentation practices and improve the quality of the refactoring documentation. Further, the refactoring motivation categories tell the opinion of developers, so it is important for managers to learn developers' opinions and feelings especially for distributed software development practices. If developers do not document, managers will not know their intention. Since software engineering is a human-centric process, it is important for managers to understand the people's intention to work on the team through their documentation.
    
    \item \textbf{Need for better tools to support the documentation of refactorings for developers at all experience levels.} Since the documentation using commit messages is usually written using natural language, and generally conveys some information about the commit they represent, writing high-quality documentation becomes vital for development and maintenance tasks \citep{treude2020beyond,plosch2014value}. As we found from RQ4 that refactoring contributors who frequently refactor the code tend to less document refactoring activities than the remaining set of developers, we plan to build a generative model based on refactoring documentation quality dimensions to automatically document refactoring properly while ensuring that there is no inconsistency between the code changes and refactoring documentation. This facilitates the automatic generation of refactoring documentation using the list of Self-Affirmed Refactoring patterns identified in \citep{alomar2019can,alomar2020toward,alomar2020howwe,alomarmining}. To demonstrate its applicability, we plan to conduct a pilot study with experts in order to assess  the framework based on the refactoring documentation quality dimensions.
    
    \item \textbf{Better understanding of refactoring best practices.} 
    Our study reveals details about developers refactoring practices. Understanding code refactoring best practices and learning from experienced developers would represent an important asset for junior developers. To push the frontier of refactoring in practice, it would be interesting to investigate the difference between top contributed and less contributed developers in terms of distributions of refactoring operations, i.e., we aim to see if any specific refactoring types are highly solicited by one group compared to the other. As previous studies have already shown, some refactoring operations tend to be more complex than others \cite{murphy2008breaking}, and so it is interesting for us to validate it in practice. Further, since our study sheds light on the driver behind refactoring performed by different groups of developers, future work can focus on recommending who should refactor the code and proposing the best solutions to refactorings.

    \item \textbf{Impact of developer's experience on software quality.} One potential research direction is to study whether developer experience is one of the factors that might contribute to the significant improvement of the quality metrics that are aligned with developer perception tagged in the commit messages. In other words, we would like to evaluate the top contributors refactoring practice against all the rest of refactoring contributors by assessing their contributions on the main internal quality attributes improvement (e.g., cohesion, coupling, and complexity). Furthermore, previous studies analyzed the impact of refactorings on structural metrics and quality attributes \cite{xing2006eclipse,bavota2015experimental,pinto2013programmers,yoshida2016revisiting}. It would be interesting to revisit such analysis while taking into account the degree of expertise of the refactoring contributors. As developers with larger experience and managerial roles have better exposure to the system's design, it is expected that their restructurings are of better quality, and this can be empirically demonstrated.

    \item \textbf{Investigating refactoring (mis)use.} With regards to the analysis of refactoring and design quality, previous studies investigated how refactorings can be responsible for introducing code smells, and so hindering the design quality \cite{tufano2015and}. It would be interesting to verify whether such unexpected results can correlate with the developer's experience. Along with hindering design quality, the misuse of refactoring can also be responsible for bugs \cite{alves2014refdistiller,bavota2012does}, and various studies have proposed testing strategies to make refactoring safer \cite{Soares2009safetytool,soares2010making}. One of our future directions is to also correlate the bug-proneness of refactorings with the degree of expertise of the contributors. It is assumed that the lack of functional knowledge may facilitate the introduction of bugs, but this is subject to empirical validation as well.

\end{itemize}

\section{Threats to Validity} \label{section:threats}


The first threat is that our analysis is restricted to only open-source, Java, Git-based repositories. However, we were still able to analyze projects that are highly varied in size, contributors, number of commits, and refactorings. Additionally, the representativeness of the dataset can be considered as a threat to this study. However, we mitigate this threat by utilizing 800 engineered projects that have also been part of a prior study on refactoring \cite{peruma2019context}. Furthermore, the projects are of varying sizes, contributors, and refactoring operations.   

The accuracy of the refactoring detection tool also poses a threat to our study. However, previous studies \cite{silva2016we,tsantalis2018accurate} report on high precision and recall scores for Refactoring Miner. However, a drawback to using Refactoring Miner is that the study is limited to Java projects. Our future work includes the use of refactoring mining tools that support other programming languages, such as RefDiff \cite{silva2017refdiff}, to expand the representativeness of our dataset.

Another potential threat to validity relates to our findings regarding counting the reported SAR patterns. Due to the large number of commit messages, we have not performed a manual validation to remove false positive commit messages. Thus, this may have an impact on our findings.

A major threat to validity is related to the calculation of experience. Obtaining the experience of each and every developer is a challenge for our study, given the volume of data in our dataset and also that experience can be subjective. Hence, we adopted a mechanism (i.e., DCR), used by prior research \cite{7972731,peruma2019context}, where we utilized project contributions as a proxy for experience. The reasoning behind the measurement assumes that the longer a developer is involved in a project, and the more they contribute to it, the more experienced they become. Such an assumption may not hold for some specific scenarios; however, since the projects in our dataset are heterogeneous in nature, our assumption holds. It is also critical to mention that we are assessing the developer's experience with respect to one project, and not looking at the broader aspect of their development expertise. In this context, developer experience indicates the degree to which they contributed to a given code base. Therefore, the measurement of developers contributions as a mean of experience holds, as our experiments our primarily focused on code changes.





\section{Conclusion} \label{sec:conclusion}
We present a study of the level of contribution of developers that apply refactorings. Prior studies use smaller samples to study similar questions; however, in our study, we have examined a more extensive and representative set of systems by comparison. Further, we explored developers practice in documenting refactoring and the distribution of refactoring operation per category.
Since we can confirm results from prior studies, we have identified a way to obtain similar results automatically. This means that it is possible to now study the impact of developer experience on a larger scale. As for the refactoring documentation and its relation to developers contribution, we found that refactoring contributors who occasionally refactor the code, tend to document refactoring more than the remaining set of developers who frequently perform refactoring.

In future work, we plan to leverage the results from this study to determine specific types of refactorings made by developers at different contribution levels. We would also like to explore ways to leverage this data to help suggest/recommend refactorings or suggest/recommend refactoring methodology based on the developers level of experience. Additionally, our empirical evidence can help future work to investigate the importance of experience in recommending who should refactoring the code.

\section*{Acknowledgments}
\label{sec:ack}
We would like to thank the authors of Refactoring Miner for publicly providing it.


\bibliography{main}%

\begin{thebibliography}{10}
\providecommand \doibase [0]{http://dx.doi.org/}%

\bibitem{codabux2013managing}
Codabux Z, Williams B. Managing technical debt: An industrial case study. In:
  {\it 2013 4th International Workshop on Managing Technical Debt (MTD)}. IEEE.
  ; 2013\string: 8--15.

\bibitem{bavota2015experimental}
Bavota G, De~Lucia A, Di~Penta M, Oliveto R, Palomba F. An experimental
  investigation on the innate relationship between quality and refactoring.
  {\it Journal of Systems and Software} 2015\string; 107.

\bibitem{fontana2012automatic}
Fontana FA, Braione P, Zanoni M. Automatic detection of bad smells in code: An
  experimental assessment.. {\it Journal of Object Technology} 2012\string;
  11(2)\string: 5--1.

\bibitem{palomba2013detecting}
Palomba F, Bavota G, Di~Penta M, Oliveto R, De~Lucia A, Poshyvanyk D. Detecting
  bad smells in source code using change history information. In:  {\it 2013
  28th IEEE/ACM International Conference on Automated Software Engineering
  (ASE)}. IEEE. ; 2013\string: 268--278.

\bibitem{palomba2016textual}
Palomba F, Panichella A, De~Lucia A, Oliveto R, Zaidman A. A textual-based
  technique for smell detection. In:  {\it 2016 IEEE 24th international
  conference on program comprehension (ICPC)}. IEEE. ; 2016\string: 1--10.

\bibitem{bavota2014recommending}
Bavota G, De~Lucia A, Marcus A, Oliveto R. Recommending refactoring operations
  in large software systems. In:  {\it Recommendation Systems in Software
  Engineering}. Springer.  2014 (pp. 387--419).

\bibitem{mkaouer2014high}
Mkaouer MW, Kessentini M, Bechikh S, Deb K, {\'O}~Cinn{\'e}ide M. High
  dimensional search-based software engineering: finding tradeoffs among 15
  objectives for automating software refactoring using NSGA-III. In:  {\it
  Proceedings of the 2014 Annual Conference on Genetic and Evolutionary
  Computation}. ACM. ; 2014\string: 1263--1270.

\bibitem{mkaouer2015many}
Mkaouer W, Kessentini M, Shaout A, et al. Many-objective software
  remodularization using NSGA-III. {\it ACM Transactions on Software
  Engineering and Methodology (TOSEM)} 2015\string; 24(3)\string: 1--45.

\bibitem{mkaouer2017robust}
Mkaouer MW, Kessentini M, Cinn{\'e}ide M{\'O}, Hayashi S, Deb K. A robust
  multi-objective approach to balance severity and importance of refactoring
  opportunities. {\it Empirical Software Engineering} 2017\string;
  22(2)\string: 894--927.

\bibitem{mkaouer2014software}
Mkaouer MW, Kessentini M, Bechikh S, {\'O}'Cinn{\'e}ide M, Deb K. Software
  refactoring under uncertainty: a robust multi-objective approach. In:  {\it
  Proceedings of the Companion Publication of the 2014 Annual Conference on
  Genetic and Evolutionary Computation}. ACM. ; 2014\string: 187--188.

\bibitem{rizzi2018support}
Rizzi L, Fontana FA, Roveda R. Support for architectural smell refactoring. In:
   {\it Proceedings of the 2nd International Workshop on Refactoring}. ACM. ;
  2018\string: 7--10.

\bibitem{terra2018jmove}
Terra R, Valente MT, Miranda S, Sales V. JMove: A novel heuristic and tool to
  detect move method refactoring opportunities. {\it Journal of Systems and
  Software} 2018\string; 138\string: 19--36.

\bibitem{de2019finding}
Oliveira dMC, Freitas D, Bonif{\'a}cio R, Pinto G, Lo D. Finding needles in a
  haystack: Leveraging co-change dependencies to recommend refactorings. {\it
  Journal of Systems and Software} 2019\string; 158\string: 110420.

\bibitem{xing2006refactoring}
Xing Z, Stroulia E. Refactoring detection based on umldiff change-facts
  queries. In:  {\it 2006 13th Working Conference on Reverse Engineering}.
  IEEE. ; 2006\string: 263--274.

\bibitem{kim2010ref}
Kim M, Gee M, Loh A, Rachatasumrit N. Ref-Finder: a refactoring reconstruction
  tool based on logic query templates. In:  {\it Proceedings of the eighteenth
  ACM SIGSOFT international symposium on Foundations of software engineering}.
  ACM. ; 2010\string: 371--372.

\bibitem{silva2017refdiff}
Silva D, Valente MT. RefDiff: detecting refactorings in version histories. In:
  {\it 2017 IEEE/ACM 14th International Conference on Mining Software
  Repositories (MSR)}. IEEE. ; 2017\string: 269--279.

\bibitem{xing2006eclipse}
Xing Z, Stroulia E. Refactoring practice: How it is and how it should be
  supported-an eclipse case study. In:  {\it 2006 22nd IEEE International
  Conference on Software Maintenance}. IEEE. ; 2006\string: 458--468.

\bibitem{pinto2013programmers}
Pinto GH, Kamei F. What programmers say about refactoring tools? an empirical
  investigation of stack overflow. In:  {\it Proceedings of the 2013 ACM
  workshop on Workshop on refactoring tools}. ACM. ; 2013\string: 33--36.

\bibitem{yoshida2016revisiting}
Yoshida N, Saika T, Choi E, Ouni A, Inoue K. Revisiting the relationship
  between code smells and refactoring. In:  {\it IEEE 24th International
  Conference on Program Comprehension (ICPC)}. IEEE. ; 2016\string: 1--4.

\bibitem{silva2016we}
Silva D, Tsantalis N, Valente MT. Why we refactor? confessions of github
  contributors. In:  {\it Proceedings of the 2016 24th ACM SIGSOFT
  International Symposium on Foundations of Software Engineering}. ACM. ;
  2016\string: 858--870.

\bibitem{peruma2018empirical}
Peruma A, Mkaouer MW, Decker MJ, Newman CD. An empirical investigation of how
  and why developers rename identifiers. In:  {\it Proceedings of the 2nd
  International Workshop on Refactoring}. ACM. ; 2018\string: 26--33.

\bibitem{alomar2020relationship}
AlOmar EA, Peruma A, Newman CD, Mkaouer MW, Ouni A. On the relationship between
  developer experience and refactoring: An exploratory study and preliminary
  results. In:  {\it Proceedings of the IEEE/ACM 42nd International Conference
  on Software Engineering Workshops}. ACM. ; 2020\string: 342--349.

\bibitem{tsantalis2013multidimensional}
Tsantalis N, Guana V, Stroulia E, Hindle A. A multidimensional empirical study
  on refactoring activity. In:  {\it Proceedings of the 2013 Conference of the
  Center for Advanced Studies on Collaborative Research}. IBM Corp. ;
  2013\string: 132--146.

\bibitem{kim2014empirical}
Kim M, Zimmermann T, Nagappan N. An empirical study of refactoringchallenges
  and benefits at microsoft. {\it IEEE Transactions on Software Engineering}
  2014\string; 40(7)\string: 633--649.

\bibitem{newman2018study}
Newman CD, Mkaouer MW, Collard ML, Maletic JI. A study on developer perception
  of transformation languages for refactoring. In:  {\it Proceedings of the 2nd
  International Workshop on Refactoring}. ACM. ; 2018\string: 34--41.

\bibitem{stroggylos2007refactoring}
Stroggylos K, Spinellis D. Refactoring--Does It Improve Software Quality?. In:
  {\it Software Quality, 2007. WoSQ'07: ICSE Workshops 2007. Fifth
  International Workshop on}. IEEE. ; 2007\string: 10--10.

\bibitem{Ratzinger:2008:RRS:1370750.1370759}
Ratzinger J, Sigmund T, Gall HC. On the Relation of Refactorings and Software
  Defect Prediction. In:  {\it Proceedings of the 2008 International Working
  Conference on Mining Software Repositories}. MSR '08. ACM. ACM; 2008; New
  York, NY, USA\string: 35--38

\bibitem{murphy2012we}
Murphy-Hill E, Parnin C, Black AP. How we refactor, and how we know it. {\it
  IEEE Transactions on Software Engineering} 2012\string; 38(1)\string: 5--18.

\bibitem{soares2013comparing}
Soares G, Gheyi R, Murphy-Hill E, Johnson B. Comparing approaches to analyze
  refactoring activity on software repositories. {\it Journal of Systems and
  Software} 2013\string; 86(4)\string: 1006--1022.

\bibitem{zhangpreliminary18}
Zhang D, Li B, Li Z, Liang P. A Preliminary Investigation of Self-Admitted
  Refactorings in Open Source Software. In: IEEE. ; 2018

\bibitem{alomar2019can}
AlOmar EA, Mkaouer MW, Ouni A. Can Refactoring Be Self-Affirmed? An Exploratory
  Study on How Developers Document Their Refactoring Activities in Commit
  Messages. In:  {\it Proceedings of the 3rd International Workshop on
  Refactoring}. IWOR ’19. IEEE. ; 2019.

\bibitem{alomar2019towards}
AlOmar EA. Towards Better Understanding Developer Perception of Refactoring.
  In:  {\it 2019 IEEE International Conference on Software Maintenance and
  Evolution (ICSME)}. IEEE. \string: 624--628.

\bibitem{alomar2020howwe}
AlOmar EA, Peruma A, Mkaouer MW, Newman CD, Ouni A, Kessentini M. How we
  refactor and how we document it? On the use of supervised machine learning
  algorithms to classify refactoring documentation. {\it Expert Systems with
  Applications} 2020\string: 114176.

\bibitem{Soares2009safetytool}
Soares G, Cavalcanti D, Gheyi R, Massoni T, Serey D, Cornélio M.
  Saferefactor-tool for checking refactoring safety.  2009.

\bibitem{Fowler:1999:RID:311424}
Fowler M, Beck K, Brant J, Opdyke W, Roberts d. {\it Refactoring: Improving the
  Design of Existing Code}.
\newblock Boston, MA, USA: Addison-Wesley Longman Publishing Co., Inc. .
\newblock 1999.

\bibitem{alomar2020toward}
AlOmar EA, Mkaouer MW, Ouni A. Toward the automatic classification of
  Self-Affirmed Refactoring. {\it Journal of Systems and Software} 2020\string:
  110821.

\bibitem{alomar2019impact}
AlOmar EA, Mkaouer MW, Ouni A, Kessentini M. On the impact of refactoring on
  the relationship between quality attributes and design metrics. In:  {\it
  2019 ACM/IEEE International Symposium on Empirical Software Engineering and
  Measurement (ESEM)}. IEEE. ; 2019\string: 1--11.

\bibitem{munaiah2017curating}
Munaiah N, Kroh S, Cabrey C, Nagappan M. Curating GitHub for engineered
  software projects. {\it Empirical Software Engineering} 2017\string;
  22(6)\string: 3219--3253.

\bibitem{peruma2019context}
{Peruma} A, {Mkaouer} MW, {Decker} MJ, {Newman} CD. Contextualizing Rename
  Decisions using Refactorings and Commit Messages. In:  {\it 2019 19th
  International Working Conference on Source Code Analysis and Manipulation
  (SCAM)}. IEEE. ; 2019\string: 74-85

\bibitem{fakhoury2019improving}
Fakhoury S, Roy D, Hassan A, Arnaoudova V. Improving source code readability:
  theory and practice. In:  {\it 2019 IEEE/ACM 27th International Conference on
  Program Comprehension (ICPC)}. IEEE. ; 2019\string: 2--12.

\bibitem{tsantalis2018accurate}
Tsantalis N, Mansouri M, Eshkevari LM, Mazinanian D, Dig D. Accurate and
  efficient refactoring detection in commit history. In:  {\it Proceedings of
  the 40th International Conference on Software Engineering}. ACM. ; 2018.

\bibitem{moser2006does}
Moser R, Sillitti A, Abrahamsson P, Succi G. Does refactoring improve
  reusability?. In:  {\it International conference on software reuse}.
  Springer. ; 2006\string: 287--297.

\bibitem{palomba2017exploratory}
Palomba F, Zaidman A, Oliveto R, De~Lucia A. An exploratory study on the
  relationship between changes and refactoring. In:  {\it 2017 IEEE/ACM 25th
  International Conference on Program Comprehension (ICPC)}. IEEE. ;
  2017\string: 176--185.

\bibitem{pantiuchina2020developers}
Pantiuchina J, Zampetti F, Scalabrino S, et al. Why developers refactor source
  code: A mining-based study. {\it ACM Transactions on Software Engineering and
  Methodology (TOSEM)} 2020\string; 29(4)\string: 1--30.

\bibitem{paixao2020behind}
Paix{\~a}o M, Uch{\^o}a A, Bibiano AC, et al. Behind the Intents: An In-depth
  Empirical Study on Software Refactoring in Modern Code Review. {\it 17th MSR}
  2020.

\bibitem{alomar2020reusability}
AlOmar EA, Rodriguez PT, Bowman J, et al. How Do Developers Refactor Code to
  Improve Code Reusability?. In:  {\it International Conference on Software and
  Software Reuse}. Springer. ; 2020\string: 261--276.

\bibitem{alomar2021xerox}
AlOmar EA, AlRubaye H, Mkaouer MW, Ouni A, Kessentini M. Refactoring Practices
  in the Context of Modern Code Review: An Industrial Case Study at Xerox. In:
  {\it 2021 IEEE/ACM 43rd International Conference on Software Engineering:
  Software Engineering in Practice (ICSE-SEIP)}. IEEE. ; 2021\string: 348--357.

\bibitem{jiang2017automatically}
Jiang S, Armaly A, McMillan C. Automatically generating commit messages from
  diffs using neural machine translation. In:  {\it 2017 32nd IEEE/ACM
  International Conference on Automated Software Engineering (ASE)}. IEEE. ;
  2017\string: 135--146.

\bibitem{Mauczka2012}
Mauczka A, Huber M, Schanes C, Schramm W, Bernhart M, Grechenig T. {\it Tracing
  Your Maintenance Work -- A Cross-Project Validation of an Automated
  Classification Dictionary for Commit Messages}\string: 301--315; Berlin,
  Heidelberg: Springer Berlin Heidelberg .
\newblock 2012

\bibitem{fu2015automated}
Fu Y, Yan M, Zhang X, Xu L, Yang D, Kymer JD. Automated classification of
  software change messages by semi-supervised Latent Dirichlet Allocation. {\it
  Information and Software Technology} 2015\string; 57\string: 369--377.

\bibitem{da2017using}
Silva~Maldonado dE, Shihab E, Tsantalis N. Using natural language processing to
  automatically detect self-admitted technical debt. {\it IEEE Transactions on
  Software Engineering} 2017\string; 43(11)\string: 1044--1062.

\bibitem{kochhar2014automatic}
Kochhar PS, Thung F, Lo D. Automatic fine-grained issue report
  reclassification. In:  {\it Engineering of Complex Computer Systems (ICECCS),
  2014 19th International Conference on}. IEEE. ; 2014\string: 126--135.

\bibitem{le2015rclinker}
Le TDB, Linares-V{\'a}squez M, Lo D, Poshyvanyk D. Rclinker: Automated linking
  of issue reports and commits leveraging rich contextual information. In:
  {\it 2015 IEEE 23rd International Conference on Program Comprehension}. IEEE.
  ; 2015\string: 36--47.

\bibitem{lane2019natural}
Lane H, Hapke H, Howard C. {\it Natural Language Processing in Action:
  Understanding, Analyzing, and Generating Text with Python}.
\newblock Manning Publications Company .
\newblock 2019.

\bibitem{Bird2002NLTKTN}
Bird S. NLTK: The Natural Language Toolkit. {\it ArXiv} 2002\string;
  cs.CL/0205028.

\bibitem{singh2013elements}
Singh R, Mangat N. {\it Elements of Survey Sampling}.
\newblock Texts in the Mathematical SciencesSpringer Netherlands .
\newblock 2013.

\bibitem{zheng2018feature}
Zheng A, Casari A. {\it Feature Engineering for Machine Learning: Principles
  and Techniques for Data Scientists}.
\newblock O'Reilly Media .
\newblock 2018.

\bibitem{manning2008introduction}
Manning C, Raghavan P, Sch{\"u}tze H. {\it Introduction to Information
  Retrieval}.
\newblock Cambridge University Press .
\newblock 2008.

\bibitem{lin2013empirical}
Lin S, Ma Y, Chen J. Empirical Evidence on Developer's Commit Activity for
  Open-Source Software Projects.. In:  {\it SEKE}. . 13. IEEE. ; 2013\string:
  455--460.

\bibitem{kowsari2019text}
Kowsari K, Jafari~Meimandi K, Heidarysafa M, Mendu S, Barnes L, Brown D. Text
  classification algorithms: A survey. {\it Information} 2019\string;
  10(4)\string: 150.

\bibitem{tan2002use}
Tan CM, Wang YF, Lee CD. The use of bigrams to enhance text categorization.
  {\it Information processing \& management} 2002\string; 38(4)\string:
  529--546.

\bibitem{cSupportVector}
Deng N, Tian Y, Zhang C. {\it Support Vector Machines: Optimization Based
  Theory, Algorithms, and Extensions}.
\newblock Chapman \& Hall/CRC Data Mining and Knowledge Discovery SeriesTaylor
  \& Francis .
\newblock 2012.

\bibitem{LIBSVM}
Chang CC, Lin CJ. LIBSVM: A Library for Support Vector Machines.  2011\string;
  2(3).
\newblock \href {\doibase 10.1145/1961189.1961199} {doi:
  10.1145/1961189.1961199}

\bibitem{CART}
Breiman L. {\it Classification and Regression Trees}.
\newblock CRC Press .
\newblock 2017.

\bibitem{Hindle:2011:ATN:1985441.1985466}
Hindle A, Ernst NA, Godfrey MW, Mylopoulos J. Automated Topic Naming to Support
  Cross-project Analysis of Software Maintenance Activities. In:  {\it
  Proceedings of the 8th Working Conference on Mining Software Repositories}.
  MSR '11. ACM. ; 2011; New York, NY, USA\string: 163--172

\bibitem{Levin:2017:BAC:3127005.3127016}
Levin S, Yehudai A. Boosting Automatic Commit Classification Into Maintenance
  Activities By Utilizing Source Code Changes. In:  {\it Proceedings of the
  13th International Conference on Predictive Models and Data Analytics in
  Software Engineering}. PROMISE. ACM. ACM; 2017; New York, NY, USA\string:
  97--106

\bibitem{honel2019importance}
H{\"o}nel S, Ericsson M, L{\"o}we W, Wingkvist A. Importance and Aptitude of
  Source code Density for Commit Classification into Maintenance Activities.
  In:  {\it The 19th IEEE International Conference on Software Quality,
  Reliability, and Security}. IEEE. ; 2019.

\bibitem{inherently}
SKlearn . 1.12. Multiclass and multilabel algorithms — scikit-learn 0.23.2
  documentation. \url{https://scikit-learn.org/stable/modules/multiclass.html};
   2007.

\bibitem{SVC}
SKlearn . sklearn.svm.SVC — scikit-learn 0.23.2 documentation.
  \url{https://scikit-learn.org/stable/modules/generated/sklearn.svm.SVC.html\#sklear.svm.SVC};
  2007.

\bibitem{dangeti2017statistics}
Dangeti P. {\it Statistics for Machine Learning}.
\newblock Packt Publishing .
\newblock 2017.

\bibitem{ProjectWebsite}
Project Website. \url{https://smilevo.github.io/self-affirmed-refactoring/}; .

\bibitem{7972731}
{Krutz} DE, {Munaiah} N, {Peruma} A, {Wiem Mkaouer} M. Who Added That
  Permission to My App? An Analysis of Developer Permission Changes in Open
  Source Android Apps. In:  {\it 2017 IEEE/ACM 4th International Conference on
  Mobile Software Engineering and Systems (MOBILESoft)}. IEEE. ; 2017\string:
  165-169

\bibitem{tukey1977exploratory}
Tukey JW. {\it Exploratory data analysis}. 2.
\newblock Reading, Mass. .
\newblock 1977.

\bibitem{murphy2008breaking}
Murphy-Hill E, Black AP. Breaking the Barriers to Successful Refactoring:
  Observations and Tools for Extract Method. In:  {\it Proceedings of the 30th
  International Conference on Software Engineering}. ICSE ’08. ACM.
  Association for Computing Machinery; 2008; New York, NY, USA\string:
  421–430

\bibitem{functional}
https://github.com/jenkinsci/subversion-plugin.
  \url{https://github.com/jenkinsci/subversion-plugin/commit/179fec8}; .

\bibitem{bugfix}
https://github.com/dustin/java-memcached-client.
  \url{https://github.com/dustin/java-memcached-client/commit/55f6911e82789e6cbad1ceccc66b1a10e609ffb5};
  .

\bibitem{codesmell}
https://github.com/ismavatar/lateralgm.
  \url{https://github.com/ismavatar/lateralgm/commit/23dfe805242599bd5abc2edf6eaa11c5379e1287};
  .

\bibitem{internal}
https://github.com/adangel/pmd.
  \url{https://github.com/adangel/pmd/commit/7f113dfe7ebc0a015c14a9c02637e39302000d01};
  .

\bibitem{external}
https://github.com/jawi/ols.
  \url{https://github.com/jawi/ols/commit/de2b95c01d3f998256abaaddaffa6fb1536f9b39};
  .

\bibitem{conover1998practical}
Conover WJ. {\it Practical nonparametric statistics}. 350.
\newblock John Wiley \& Sons .
\newblock 1998.

\bibitem{alomar2021SLM}
AlOmar EA, Mkaouer MW, Newman C, Ouni A. On Preserving the Behavior in Software
  Refactoring: A Systematic Mapping Study. {\it Information and Software
  Technology} 2021.

\bibitem{treude2020beyond}
Treude C, Middleton J, Atapattu T. Beyond accuracy: assessing software
  documentation quality. In:  {\it Proceedings of the 28th ACM Joint Meeting on
  European Software Engineering Conference and Symposium on the Foundations of
  Software Engineering}. ACM. ; 2020\string: 1509--1512.

\bibitem{plosch2014value}
Pl{\"o}sch R, Dautovic A, Saft M. The value of software documentation quality.
  In:  {\it 2014 14th International Conference on Quality Software}. IEEE. ;
  2014\string: 333--342.

\bibitem{alomarmining}
AlOmar EA, Mkaouer MW, Ouni A. Mining and Managing Big Data Refactoring for
  Design Improvement: Are We There Yet?. {\it Knowledge Management in the
  Development of Data-Intensive Systems}\string: 127--140.

\bibitem{tufano2015and}
Tufano M, Palomba F, Bavota G, et al. When and why your code starts to smell
  bad. In:  {\it 2015 IEEE/ACM 37th IEEE International Conference on Software
  Engineering}. . 1. IEEE. ; 2015\string: 403--414.

\bibitem{alves2014refdistiller}
Alves EL, Song M, Kim M. RefDistiller: a refactoring aware code review tool for
  inspecting manual refactoring edits. In:  {\it Proceedings of the 22nd ACM
  SIGSOFT International Symposium on Foundations of Software Engineering}. ACM.
  ; 2014\string: 751--754.

\bibitem{bavota2012does}
Bavota G, De~Carluccio B, De~Lucia A, Di~Penta M, Oliveto R, Strollo O. When
  does a refactoring induce bugs? an empirical study. In:  {\it IEEE 12th
  International Working Conference on Source Code Analysis and Manipulation}.
  IEEE. ; 2012\string: 104--113.

\bibitem{soares2010making}
Soares G, Gheyi R, Serey D, Massoni T. Making program refactoring safer. {\it
  IEEE software} 2010\string; 27(4)\string: 52--57.

\end{thebibliography}



\end{document}